\documentclass[journal]{IEEEtran}
\IEEEoverridecommandlockouts
\usepackage{bm}
\usepackage{cite}
\usepackage{amsmath,amssymb,amsfonts,steinmetz}
\usepackage{graphicx}
\usepackage{textcomp}
\usepackage{xcolor}
\def\BibTeX{{\rm B\kern-.05em{\sc i\kern-.025em b}\kern-.08em
    T\kern-.1667em\lower.7ex\hbox{E}\kern-.125emX}}

\usepackage{pgfplots}
\usepackage[font=footnotesize]{caption}
\usepackage{subcaption}
\usepackage{bbm}
\usepackage{tikz}
\usetikzlibrary{calc}
\makeatletter
\newcommand{\gettikzxy}[3]{%
  \tikz@scan@one@point\pgfutil@firstofone#1\relax
  \edef#2{\the\pgf@x}%
  \edef#3{\the\pgf@y}%
}
\makeatother
\usepackage[draft]{todonotes}   % notes showed
\usepackage{upgreek} %use for mathbf symbols like \bm{theta} = \mathbf{\uptheta}
\usepackage{seqsplit}
\usepackage{hyperref}
\usepackage{balance}
\usepackage{acronym}

\usepackage[]{algorithm}
\usepackage[noend]{algpseudocode}

\makeatletter
\newcommand*\rel@kern[1]{\kern#1\dimexpr\macc@kerna}
\newcommand*\widebar[1]{%
  \begingroup
  \def\mathaccent##1##2{%
    \rel@kern{0.8}%
    \overline{\rel@kern{-0.8}\macc@nucleus\rel@kern{0.2}}%
    \rel@kern{-0.2}%
  }%
  \macc@depth\@ne
  \let\math@bgroup\@empty \let\math@egroup\macc@set@skewchar
  \mathsurround\z@ \frozen@everymath{\mathgroup\macc@group\relax}%
  \macc@set@skewchar\relax
  \let\mathaccentV\macc@nested@a
  \macc@nested@a\relax111{#1}%
  \endgroup
}
\makeatother

\newcommand{\herm}{{\textrm{H}}}
\newcommand{\diag}{\mathrm{diag}}

\newcommand{\vind}[1]{\boldsymbol{#1}}
\newcommand{\matInd}[1]{\boldsymbol{#1}}

\newcommand{\mr}[1]{\mathrm{#1}}

\newcommand{\numSubCarriers}{N}

\newcommand{\deltaf}{\Delta_f}

\newcommand{\chParMark}{\mr{ch}}
\newcommand{\fisherInfMark}{\matInd{F}}
\newcommand{\positionOrientationMark}{\mr{po}}
\newcommand{\fisherInfpo}{\fisherInfMark_{\positionOrientationMark}}
\newcommand{\fisherInfch}{\fisherInfMark_{\chParMark}}
\newcommand{\parameterMark}{\vind{\zeta}}
\newcommand{\parameterCh}{\parameterMark_{\chParMark}}

\newcommand{\parameterPo}{\parameterMark_{\positionOrientationMark}}

\newcommand{\jacob}{\matInd{J}}
\newcommand{\fisherInfoch}{\fisherInfMark_{\chParMark}}
\newcommand{\realSet}{\mathbb{R}}
\newcommand{\rond}{\partial}

\newcommand{\realPart}{\operatorname{Re}}
\newcommand{\vecNorm}[1]{{\left\lVert#1\right\rVert}}

\DeclareMathOperator{\atant}{\mr{atan2}}
\newcommand{\meanSig}{\bm{M}}

\newcommand{\Tsym}{ T_{\rm{sym}} }
\newcommand{\Tcp}{ T_{\rm{cp}} }

\newcommand{\diagg}[1]{ {\mathrm{diag}}\left(#1\right)  }

\newcommand{\vub}{v_{\mathrm{b}}}
\newcommand{\vurk}{v_{\mathrm{r}}}
\newcommand{\murk}{\nu_{\mathrm{r}}}
\newcommand{\sll}{s_{\ell}}
\newcommand{\xnl}{ x_{n,\ell} }
\newcommand{\xnltilde}{ \widetilde{x}_{n,\ell} }
\newcommand{\thn}[1]{ {#1\text{th} } }
\newcommand{\rect}[1]{ { \rm{rect} }\left(#1\right) }

\newcommand{\Tso}{ T_{\mathrm{o}} }
\newcommand{\complexset}[2]{ \mathbb{C}^{#1 \times #2}  }

\newcommand{\fc}{ f_c }
\newcommand{\stilde}{ s_{\mathrm{u}} }
\newcommand{\ytilde}{ y_{\mathrm{ur}} }
\newcommand{\gr}{g_{\mathrm{r}}}
\newcommand{\grtilde}{\widetilde{g}_{\mathrm{r}}}

\newcommand{\Taubr}{ \bm{\tau}_{\rm{br}} }
\newcommand{\Tauru}{ \bm{\tau}_{\rm{ru}} }
\newcommand{\Taur}{ \bm{\tau}_{\rm{r}} }
\newcommand{\Tauv}{ \bm{\tau} }

\newcommand{\taumbr}{ [\bm{\tau}_{\rm{br}}]_m }

\newcommand{\taumru}{[\bm{\tau}_{\rm{ru}}]_m}

\newcommand{\taumgell}{ [\bm{\tau}_{\rm{r},\ell}]_m }

\newcommand{\taumin}{ \tau_{\rm{min}} }

\newcommand{\ylkm}{[\widetilde{\bm{Y}}_{\mathrm{r}}]_{k,\ell} }

\newcommand{\deltat}{ \Delta_t }

\newcommand{\thetab}{ \bm{\theta} }
\newcommand{\phib}{ \bm{\phi} }

\newcommand{\aabig}{ \bm{A} }
\newcommand{\wideband}{{\rm{w}}}
\newcommand{\aabigw}{ \aabig_{\wideband} }

\newcommand{\ccbig}{ \bm{C} }
\newcommand{\ccbigw}{ \ccbig_{\wideband} }

\newcommand{\pbb}{ \bm{p}_{\mathrm{b}} }
\newcommand{\pbrm}{ \bm{p}_{\mathrm{r},m} }
\newcommand{\pb}{ \bm{p} }
\newcommand{\norm}[1]{\left\lVert#1\right\rVert}

\acrodef{cdf}[CDF]{cumulative distribution function}
\acrodef{bs}[BS]{base station}
\acrodefplural{bs}[BSs]{base station}
\acrodef{ue}[UE]{user equipment}
\acrodef{los}[LOS]{line-of-sight}
\acrodef{aoa}[AoA]{angle-of-arrival}
\acrodefplural{aoa}[AoA]{angles-of-arrival}
\acrodef{aod}[AoD]{angle-of-departure}
\acrodef{toa}[ToA]{time-of-arrival}
\acrodef{tdoa}[TDoA]{time-difference-of-arrival}
\acrodef{ris}[RIS]{reconfigurable intelligent surface}
\acrodefplural{ris}[RISs]{reconfigurable intelligent surfaces}
\acrodef{tx}[Tx]{transmitter}
\acrodef{rx}[Rx]{receiver}
\acrodefplural{rx}[Rxs]{receivers}
\acrodef{crb}[CRB]{Cram\'er-Rao lower bound}
\acrodef{rss}[RSS]{received signal strength}
\acrodef{los}[LOS]{line-of-sight}
\acrodef{nlos}[NLOS]{non line-of-sight}
\acrodef{dft}[DFT]{discrete Fourier transform}
\acrodef{fft}[FFT]{fast Fourier transform}
\acrodef{fim}[FIM]{Fisher information matrix}
\acrodef{upa}[UPA]{uniform planar array}
\acrodef{peb}[PEB]{position error bound}
\acrodef{snr}[SNR]{signal-to-noise ratio}
\acrodef{sre}[SRE]{smart radio environment}
\acrodefplural{sre}[SRE]{smart radio environments}
\acrodef{mimo}[MIMO]{multiple-input  multiple-output}
\acrodef{rfid}[RFID]{radio-frequency identification}
\acrodef{ofdm}[OFDM]{orthogonal frequency-division multiplexing}
\acrodef{rhs}[RHS]{right-hand-side}
\acrodef{lhs}[LHS]{left-hand-side}
\acrodef{ifft}[IFFT]{inverse fast Fourier transform}

\acrodef{bs}[BS]{base station}
\acrodef{ue}[UE]{user equipment}
\acrodef{los}[LOS]{line-of-sight}
\acrodef{aoa}[AoA]{angle-of-arrival}
\acrodef{aod}[AoD]{angle-of-departure}
\acrodef{toa}[ToA]{time-of-arrival}
\acrodef{tdoa}[TDoA]{time-difference-of-arrival}
\acrodef{ris}[RIS]{reconfigurable intelligent surface}
\acrodefplural{ris}[RISs]{reconfigurable intelligent surfaces}
\acrodef{tx}[Tx]{transmitter}
\acrodef{rx}[Rx]{receiver}
\acrodefplural{rx}[Rxs]{receivers}
\acrodef{crb}[CRB]{Cram\'er-Rao lower bounds}
\acrodef{rss}[RSS]{received signal strength}
\acrodef{los}[LOS]{line-of-sight}
\acrodef{nlos}[NLOS]{non-line-of-sight}
\acrodef{dft}[DFT]{discrete Fourier transform}
\acrodef{fft}[FFT]{fast Fourier transform}
\acrodef{fim}[FIM]{Fisher information matrix}
\acrodef{upa}[UPA]{uniform planar array}
\acrodefplural{upa}[UPAs]{uniform planar arrays}
\acrodef{peb}[PEB]{position error bound}
\acrodef{snr}[SNR]{signal-to-noise ratio}
\acrodef{sre}[SRE]{smart radio environment}
\acrodefplural{sre}[SRE]{smart radio environments}
\acrodef{mimo}[MIMO]{multiple-input  multiple-output}
\acrodef{miso}[MISO]{multiple-input  single-output}
\acrodef{rfid}[RFID]{radio-frequency identification}
\acrodef{siso}[SISO]{single-input single-output}
\acrodef{ici}[ICI]{inter-carrier interference}
\acrodef{iid}[iid]{independent and identically distributed}
\acrodef{qos}[QoS]{quality of service}
\acrodef{gps}[GPS]{global positioning system}
\acrodef{rhs}[RHS]{right hand side}
\acrodef{lhs}[LHS]{left hand side}

\newcommand{\rev}[1]{\textcolor{black}{#1}} %

\DeclareRobustCommand{\vectt}[1]{\bm{#1}}
\def\app#1#2{%
  \mathrel{%
    \setbox0=\hbox{$#1\sim$}%
    \setbox2=\hbox{%
      \rlap{\hbox{$#1\propto$}}%
      \lower1.1\ht0\box0%
    }%
    \raise0.25\ht2\box2%
  }%
}

\allowdisplaybreaks
\begin{document}

\title{RIS-Enabled SISO Localization under User Mobility and Spatial-Wideband Effects}

\author{\IEEEauthorblockN{Kamran Keykhosravi,~\IEEEmembership{Member,~IEEE,} Musa Furkan Keskin,~\IEEEmembership{Member,~IEEE,} Gonzalo Seco-Granados,~\IEEEmembership{Senior~Member,~IEEE,} Petar Popovski,~\IEEEmembership{Fellow,~IEEE,} and Henk Wymeersch,~\IEEEmembership{Senior~Member,~IEEE.}}\\

	\thanks{Parts of this paper  have been  presented at the IEEE International Conference on Communications 2021 \cite{keykhosravi2020siso}. }
\thanks{ This work was supported, in part, by the Swedish Research Council under grant 2018-03701, the EU H2020 RISE-6G project under grant 101017011, Alice and Lars Erik Landahls fond under grant 90211117, Chalmers research fond under grant 90211102, the Spanish Ministry of Science and Innovation PID2020-118984GB-I00 and by the Catalan ICREA Academia Programme.}
\thanks{K. Keykhosravi, M.F. Keskin, and H. Wymeersch are with the Department of Electrical Engineering,
	Chalmers University of Technology, Gothenburg 41296, Sweden (e-mail: kamrank@chalmers.se); G. Seco-Granados is with the Department of Telecommunications and
Systems Engineering, Universitat Auton\`{o}ma de Barcelona, Spain; and Petar Popovski is with the Department of Electronic Systems, Aalborg University, Denmark. }}

\maketitle

\begin{abstract}
Reconfigurable intelligent surface (RIS) is a promising technological enabler for the 6th generation (6G) of wireless systems with applications in localization and communication. In this paper, we consider the problem of positioning a single-antenna user in 3D space based on the received signal from a single-antenna base station and reflected signal from an RIS by taking into account the mobility of the user and spatial-wideband (WB) effects. To do so, we first derive the spatial-WB channel model under the far-field assumption, for orthogonal frequency-division multiplexing signal transmission with the user having a constant velocity. We derive the Cram\'er Rao bounds to serve as a benchmark. Furthermore, we devise a low-complexity estimator that attains the bounds in high signal-to-noise ratios. Our estimator neglects the spatial-WB effects and deals with the user mobility by estimating the radial velocities and compensating for their effects in an iterative fashion. We show that the spatial-WB effects can degrade the localization accuracy for large RIS sizes and large signal bandwidths as the direction of arrival or departure deviate from the RIS normal. In particular, for a 64 $\times$ 64 RIS, the proposed estimator is resilient against the spatial-WB effects up to 140 MHz bandwidth. Regarding user mobility, our results suggest that the velocity of the user influences neither the bounds nor the accuracy of our estimator. Specifically, we observe that the state of the user with a high speed (42 m/s) can be estimated virtually with the same accuracy as a static user.   

\end{abstract}


\section{Introduction}
Estimation of user location has become increasingly crucial in today's networking technology with applications in autonomous driving, navigation, data transmission, augmented reality, etc. \cite{bourdoux20206g}. Satellite localization systems such as the \ac{gps} have the downside that they do not function properly in indoor scenarios, urban canyons, or tunnels. As a complementary approach, cellular localization can be used, where the user state is estimated based on the radio signals interchanged between the \ac{bs} and the user. Provisioning of cellular localization was stirred by the governmental authorities demanding that the operators should provide the location of the \ac{ue} upon receiving emergency calls. In 4G wireless systems, the \ac{ue} location and clock bias are estimated by calculating \ac{tdoa} between the \ac{ue} and four synchronized \acp{bs} \cite{3gpp.36.855}. In 5G, the multi-antenna structure of BSs and \acp{ue} allowed networks to also use the angles of arrival and departure for localization, enabling positioning with one \ac{bs} under rich multipath conditions \cite{Shahmansoori18TWC}. In this work, we show that the next generation, 6G, can benefit from the new technological enablers, such as \acp{ris}, to estimate the \ac{ue} position, clock bias, and velocity, even for \ac{siso} wireless links.

\acp{ris} are thin surfaces made of sub-wavelength unit cells, whose response to the impinging electromagnetic wave can be controlled \cite{chunhua_mag21}. Recently, a great deal of attention has been drawn to \acp{ris} as one of the foremost technological enablers of the next generation of wireless systems (see \cite{marco_smart} for an excellent literature review). \acp{ris} introduce a new paradigm in wireless systems since they enable the optimization of the channel to maximize the \ac{qos} \cite{2019Basar,joint_BS_RIS_BF_TCOM_2021_Poor,RIS_WCM_2021,RIS_EE_TWC_2019}.  In a communication system, where the \ac{ris} response can be optimized to improve the \ac{snr} and the spectral efficiency  at the \ac{ue} site, the main challenges pertain to \rev{path loss modeling \cite{RIS_PL_Model_TWC_2021}}, estimation of the propagation channels to/from the RIS elements \cite{swindlehurst2021channel,araujo_jstsp21,CE_IRS_TWC_2020,RIS_WCM_2021,cascadedCE_RIS_WCL_2020}, as well as the use of this estimate to employ optimized configuration of the RIS elements \rev{\cite{IRS_OFDM_TCOM_2020,JointActivePassiveRIS_TWC_2019}}.
 In radio localization, \acp{ris} can provide a strong and controllable \ac{nlos} signal path, as well as an extra location reference. 

Many works have studied the benefits of RISs in radio localization through deriving \ac{crb} and/or by designing estimation algorithms that use the reflected signal from the RIS to improve or enable \ac{ue} localization \cite{elzanaty2020reconfigurable,dardari_spawk,rahal2021ris,keykhosravi2021semi,zhang2020towards,habo_rss,sha_18,Yiming_ICC21,cramer_juan,haobo_2020,abu2020near,nearFieldRIS_LOSBlock_2022,LOS_NLOS_NearField_2021,RIS_loc_2021_TWC}. In \cite{elzanaty2020reconfigurable}, the \ac{crb} on the location and orientation of the \ac{ue} have been derived for a \ac{mimo} system equipped with an \ac{ris}, where considerable improvements in estimation accuracy have been observed because of the \ac{ris}. It has been shown that 3D localization is possible in an \ac{ris}-equipped SISO system \cite{keykhosravi2020siso,rahal2021ris}. Furthermore, in \cite{dardari_spawk,LOS_NLOS_NearField_2021}, SISO localization is performed with the help of a stripe-like \ac{ris} with blocked \ac{los} path even when the path from \ac{ris} to \ac{ue} is obstructed severely. Localization in the near-field of the \ac{ris} through analyzing the \ac{crb} has been studied in \cite{sha_18} for infinite phase resolution and in \cite{cramer_juan} for limited one. \rev{Moreover, in \cite{nearFieldRIS_LOSBlock_2022}, an uplink near-field localization algorithm is proposed for RIS-aided scenarios with \ac{los} blockage. To estimate and counteract such blockages, a joint beam training and positioning method is developed in \cite{RIS_loc_2021_TWC} in multi-RIS assisted mmWave communications.}  

Most of the aforementioned works consider quasi-static channels, where the movement of the \ac{ue} during pilot transmission is negligible. While the effect of \ac{ue} mobility \rev{has been unexplored} in RIS-based localization literature, a number of works consider \ac{ue} mobility for RIS-aided communication systems \cite{Matthiesen_continuous,basar2019reconfigurable,refractHighMob_TWC_2021,sun_wcl_doppler,RIS_MC_Doppler_TVT_2022,HST_WCL_2022,Doppler_RIS_Entropy_2022,PredictableDoppler_VTC_2021,DopplerMitigation_RIS_WCL_2022}. A continuous-time model for RIS-aided satellite communication has been derived in \cite{Matthiesen_continuous}, where the movement of the satellite has been taken into consideration in optimization of the RIS phase shifts. \rev{Similarly, \cite{PredictableDoppler_VTC_2021} investigates RIS phase shift design to simultaneously minimize the delay and Doppler spread and maximize the SNR in RIS-aided high-mobility vehicular communications under predictable \ac{ue} mobility.} In \cite{basar2019reconfigurable}, it has been shown that the  multipath fading effect caused by \ac{ue} movement can be mitigated in an RIS-aided scenario. In \cite{refractHighMob_TWC_2021}, the authors presented a transmission protocol for channel estimation in a high-mobility scenario, where \rev{an intelligent \textit{refracting} surface} is mounted on the car. \rev{Following a similar approach, the study in \cite{DopplerMitigation_RIS_WCL_2022} proposes a two-stage transmission protocol and channel/Doppler estimation method in high-mobility RIS-aided scenarios, complemented by the design of RIS phase shifts to mitigate the RIS-induced Doppler effect.} Two channel estimation schemes for an RIS-aided communication system have been proposed in \cite{sun_wcl_doppler}, considering Doppler effects. \rev{Moreover, the study in \cite{RIS_MC_Doppler_TVT_2022} models doubly-selective high-mobility Rician channels in RIS-aided unmanned aerial vehicle (UAV) communications by including the Doppler effect, and deals with minimum mean squared error (MMSE) channel
estimation and RIS phase shift optimization. Furthermore, a deep reinforcement learning-based method is proposed in \cite{HST_WCL_2022} to jointly design \ac{bs} beamforming and RIS phase shifts for RIS-assisted mmWave high-speed railway networks.} 

The spatial-wideband (WB) effect refers to the change of an array's response (spatial steering vector) due to the change in frequency within the signal bandwidth \cite{wang2018spatial},\rev{\cite{RIS_overview_2022}}. This can cause the beam-squint effect in far-field \cite{cai_squint16,hybrid_sac} and the misfocus effect in near-field \cite{infocus}. The spatial-WB effect has been studied for the case of massive \ac{mimo} (see e.g. \cite{wang2018spatial,cai_squint16,hybrid_sac}) and also recently for RISs \cite{dovelosintelligent,face_squint,chen2021beam}. In \cite{wang2018spatial}, the authors develop a spatial-WB channel model, and tailored a channel estimation algorithm based on it. In \cite{cai_squint16}, the effects of beam-squint have been analyzed and compensated for in designing analog codebooks. A channel estimation algorithm for a spatial-WB RIS-aided communication system has been proposed in \cite{face_squint}. Several RIS phase shift designs have been proposed in \cite{chen2021beam} to maximize information rate in the presence of the beam-squint effect. To the best of our knowledge, the combined contribution of \ac{ue} mobility and spatial-WB effect have not yet been studied in the context of RIS-localization.

This paper extends our conference contribution in \cite{keykhosravi2020siso}, where it was shown that in a  \ac{siso} system equipped with a single RIS, 3D \ac{ue} localization and synchronization is possible. In this paper, we define and study the problem of RIS-aided SISO localization under spatial-WB effects and user mobility. The \rev{main} contributions of this paper \rev{can be summarized} as follows.
\begin{itemize}
    \item \rev{For the first time in the literature, we investigate the problem of single-snapshot RIS-aided SISO 3D localization and synchronization under \ac{ue} mobility and spatial-WB effects.}
    
    \item We develop a geometric channel model for \ac{ofdm} signal propagation under the far-field assumption, by \rev{explicitly} taking into account \ac{ue} mobility and spatial-WB effects. \rev{Unlike the studies on RIS-aided communications with \ac{ue} mobility \cite{Matthiesen_continuous,basar2019reconfigurable,refractHighMob_TWC_2021,sun_wcl_doppler,RIS_MC_Doppler_TVT_2022,HST_WCL_2022,Doppler_RIS_Entropy_2022,PredictableDoppler_VTC_2021,DopplerMitigation_RIS_WCL_2022}, the developed model formulates the \ac{los} (i.e., BS-to-UE) and \ac{nlos} (i.e., BS-to-\ac{ris}-to-UE) channels as a function of individual geometric parameters consisting of delays, Doppler shifts, and \acp{aod} in azimuth and elevation. In addition, unlike the existing literature on RIS-aided localization \cite{elzanaty2020reconfigurable,dardari_spawk,rahal2021ris,keykhosravi2021semi,zhang2020towards,habo_rss,sha_18,Yiming_ICC21,cramer_juan,haobo_2020,abu2020near,nearFieldRIS_LOSBlock_2022,LOS_NLOS_NearField_2021,RIS_loc_2021_TWC}, we incorporate Doppler shift into our model.}
    
    \item We design a low-complexity algorithm \rev{for joint localization and synchronization of \ac{ue}, accompanied by time-orthogonal RIS phase profile design to combat interpath interference. First, we estimate the channel gain, delay and Doppler of the \ac{los} path, and subtract its effect from the received signal. Based on the resulting \ac{los}-interference-eliminated signal, we then estimate the parameters of the \ac{nlos} path, involving the delay, Doppler and \ac{aod} from the RIS to \ac{ue}. In the final stage, 3D position and clock bias of the \ac{ue} are computed using the estimated geometric channel parameters. The proposed algorithm attains} the theoretical bounds at high SNRs when the spatial-WB effects are negligible. 
    
    %for estimating the \ac{ue} state that can attain the theoretical bounds at high SNRs when the spatial-WB effects are negligible. Our estimator is designed based on a spatial-narrowband (NB) model, but it takes into consideration the \ac{ue} mobility. 
    \item We study the influence of  \ac{ue} mobility, spatial-WB effects, and the presence of scatterers on the estimation error through extensive simulation of the estimator and evaluation of the \ac{crb}, considering directional and random \ac{ris} phase profiles.
\end{itemize}
Our results suggest that in terms of fundamental bounds, neither \ac{ue} mobility nor spatial-WB effects influence the estimation accuracy. However, in terms of the accuracy of the estimator (designed based on the spatial-narrowband (NB) model), the spatial-WB effects reduce the position accuracy for large sizes of the RIS and large bandwidths when the angle between the direction of arrival or departure and the \ac{ris} normal is large. The performance of our estimator is not affected by the \ac{ue} speed.

\vspace{.5cm}
\paragraph*{Organization} The remainder of the paper is organized as follows. In Section\,\ref{sec:systemModChannelModel}, we present the system setup and derive the channel model in Section\,\ref{sec:extended-channel-models}. The \ac{ris} phase profile design is presented in Section\,\ref{sec:RisPhaseDesign}. The estimator is described in Section\,\ref{sec_estimator} through a number of separate algorithms. In Section\,\ref{sec:simulationResults}, we calculate the estimation errors through simulation and compare them with the \ac{crb} for an example of system parameters. Finally, Section\,\ref{sec:conclusion} concludes the paper.

\vspace{.5cm}
\paragraph*{Notation}
We represent  vectors by bold-face lowercase letters (e.g., $\bm{x}$) and  matrices by  bold-face uppercase ones (e.g., $\bm{X}$). The $n$th element of the vector $\bm{x}$ is shown by $[\bm{x}]_n$ and with $[\bm{X}]_{m,n}$ we indicate the element on the $m$th row and the $n$th column of matrix $\bm{X}$. Furthermore, $[\bm{X}]_{:,n}$ ($[\bm{X}]_{n,:}$) denote the $n$th column (row) of matrix $\bm{X}$. The subindex $m:n$ indicates all the elements between (and including) $m$ and $n$. The Kronecker product is shown by $\otimes$ and the Hadamard product by $\odot$. The real and imaginary parts of the complex number $x$ are shown by $\Re(x)$ and $\Im(x)$, respectively. The matrix vectorization operator is indicated by $\mathrm{vec}(\cdot)$. The vector $\bm{1}_L$ indicates the vector of length $L$, all of whose elements are one.

\section{System and channel model}\label{sec:systemModChannelModel}

\subsection{System setup}\label{sec:systemSetup}
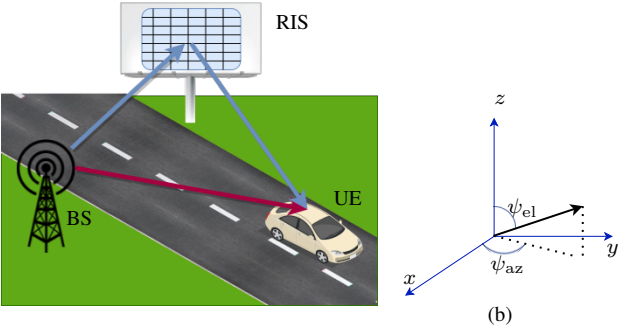
\begin{figure}
    \centering
    \begin{tikzpicture}
    \node (image) [anchor=south west]{\includegraphics[width=5cm]{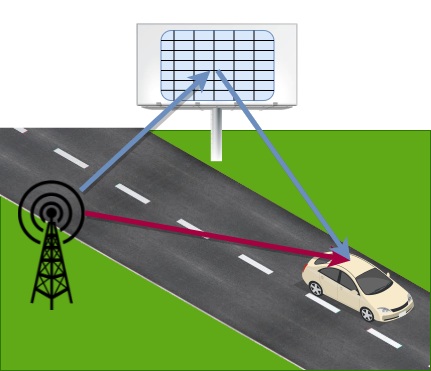}};
    \gettikzxy{(image.north east)}{\ix}{\iy};
    %\draw [help lines] (0,0) grid (\ix,\iy);
    \node at (1.1/5*\ix,1/3.5*\iy){\footnotesize BS};
    \node at (3.8/5*\ix,3/3.5*\iy){\footnotesize RIS};
    \node at (4.5/5*\ix,1.2/3.5*\iy){\footnotesize UE};
    %ue coordinates
    %\node at (1/5*\ix,.15/3.5*\iy){\footnotesize$\ueCoordinateVector{2}$};
    %\node at (.5/5*\ix,1/3.5*\iy){\footnotesize$\ueCoordinateVector{1}$};
    %\node at (1.3/5*\ix,1.7/3.5*\iy){\footnotesize$\ueCoordinateVector{3}$};
    %RIS coordinates
    %\node at (4.3/5*\ix,3.3/3.5*\iy){\footnotesize$\vind{v}_3$};
    %\node at (3.5/5*\ix,2.7/3.5*\iy){\footnotesize$\vind{v}_1$};
    %\node at (4/5*\ix,1.4/3.5*\iy){\footnotesize$\vind{v}_2$};
    %positions
    %\node at (4.5/5*\ix,2.1/3.5*\iy){\footnotesize$\bm{p}_{\mathrm{r}}$};
    %\node at (1.5/5*\ix,.9/3.5*\iy){\footnotesize$\bm{p}$};
    %\node at (.7/5*\ix,2.7/3.5*\iy){\footnotesize$\bsPosition$};
    %\node at (3.1/5*\ix,0/3.5*\iy){\footnotesize (a)};
    % right image
    \newcommand\w{3}
    \node (image2) at (\ix,0) [anchor=south west]{\includegraphics[width=\w cm]{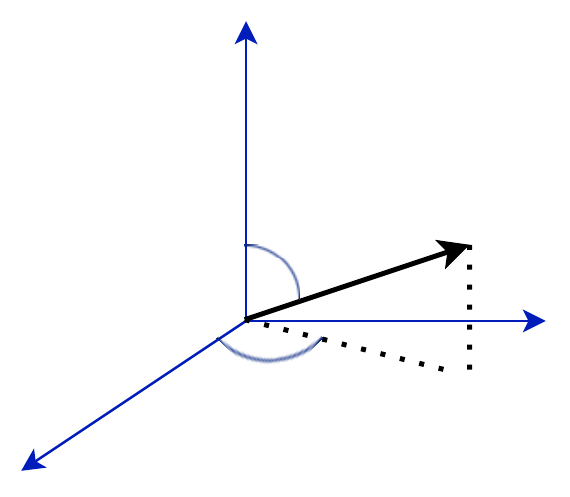}};
    \gettikzxy{(image2.north east)}{\ixt}{\iyt};
    %\draw [help lines] (\ix,0) grid (\ixt,\iyt);
    %coordinates
    \node at (\ix+.3/3*\w cm,.5/3*\iyt){\footnotesize$x$};
    \node at (\ix+3/3*\w cm,.9/3*\iyt){\footnotesize$y$};
    \node at (\ix+1.5/3*\w cm,3/3*\iyt){\footnotesize$z$};
    %angels
    \node at (\ix+1.6/3*\w cm,.7/3*\iyt){\footnotesize$\psi_{\mr{az}}$};
    \node at (\ix+1.8/3*\w cm,1.5/3*\iyt){\footnotesize$\psi_{\mr{el}}$};
    \node at (\ix+.5*\w cm,0/3.5*\iy){\footnotesize (b)};
    \end{tikzpicture}
    \caption{(a): System setup, (b): Elevation and azimuth angles of a generic vector. }
    \label{fig:setup}
\end{figure}

We consider a wireless system with a single-antenna transmitter, one \ac{ris}, and a single-antenna \ac{ue} as shown in Fig.\,\ref{fig:setup}(a).  We indicate the position of the \ac{bs} and the \ac{ris} center by $\bm{p}_{\mathrm{b}}\in \mathrm{R}^3$ and $\bm{p}_{\mathrm{r}}\in \mathrm{R}^3$ according to some general coordinate system. The values of $\bm{p}_{\mathrm{b}}$ and $\bm{p}_{\mathrm{r}}$ as well as the orientation  of the \ac{ris}  are assumed to be known. \rev{Additionally, we assume that the \ac{ue} is not time-synchronized to the \ac{bs}, leading to an unknown clock bias $\Delta_t\in \mathrm{R}$ at the \ac{ue} with respect to the \ac{bs}.} In addition to the \ac{ue}'s position ($\bm{p}\in \mathrm{R}^3$) \rev{and} clock bias $\Delta_t$, \rev{its} velocity ($\bm{v}\in \mathrm{R}^3$) \rev{is} unknown and to be estimated. The \ac{ris} is a \ac{upa} with $M = M_1\times M_2$ elements. The element in the $r$th row ($r \in \{0, \dots, M_{1}-1\}$) and $s$th column ($s \in \{0, \dots, M_{2}-1\}$) has the position $\bm{q}_{r,s} = [dr-d(M_{1}-1)/2),0,ds-d (M_{2}-1)/2 ]$ in the local coordinate system of \ac{ris}, with $d$ being the spacing between the elements. The phase profile matrix of the \ac{ris} at time $\ell$ is shown by $\bm{\Gamma}_{\ell}\in\mathbb{C}^{M_1\times M_2}$, where $[\bm{\Gamma}_\ell]_{r,s}$ indicates the phase shift applied to the impinging signal via the \ac{ris} element in the $r$th row and $s$th column.

\subsection{Geometric relations} %Between system and geometric channel parameters}
We introduce $\vub$ and $\vurk$ as the \ac{ue}'s radial velocity (Doppler) along UE-BS and UE-RIS directions, respectively, and are given by 
 \begin{align}
     \vub &= \bm{v}^{\top} (\bm{p}_{\mathrm{b}}-\bm{p})/\Vert\bm{p}_{\mathrm{b}}-\bm{p}\Vert \label{eq:vub}\\
     \vurk &= \bm{v}^{\top} (\bm{p}_{\mathrm{r}}-\bm{p})/\Vert\bm{p}_{\mathrm{r}}-\bm{p}\Vert.\label{eq:vur}
 \end{align}
 In addition, $\tau_{\mathrm{b}}$  and $\tau_{\mathrm{r}}$ represent, respectively, the delays of the direct and the reflected paths
  \begin{align}
     \tau_{\mathrm{b}} &= \frac{\Vert\bm{p}_{\mathrm{b}}-\bm{p}\Vert}{c}+\Delta_t\label{eq:taub}\\
     \tau_{\mathrm{r}}&=\frac{\Vert\bm{p}_{\mathrm{b}}-\bm{p}_{\mathrm{r}}\Vert+\Vert\bm{p}_{\mathrm{r}}-\bm{p}\Vert}{c}+\Delta_t,\label{eq:taur}
 \end{align}
  where $\Delta_t$ is the clock bias and $c$ is the light speed. 
The \ac{aod} from the \ac{ris} to the \ac{ue} is indicated by $\bm{\phi}$, which corresponds to the direction of the vector $\bm{s}$ from the \ac{ris} to the \ac{ue} in the local coordinate system of the \ac{ris}, i.e., $\bm{s} = \bm{R} (\bm{p}-\bm{p}_{\mathrm{r}})$, where $\bm{R}$ is a rotation matrix that maps the global frame of reference to the \ac{ris} local coordinate system. More specifically, we have
\begin{align}
[\bm{\phi}]_{\mathrm{az}}
&=\atant\left( [\bm{s}]_{2}, [\bm{s}]_{1} \right)\label{eq:phiAz}\\
[\bm{\phi}]_{\mathrm{el}} &=\arccos \left(\frac{[\bm{s}]_{3}}{\vecNorm{\bm{p}-\bm{p}_{\mathrm{r}}}}\right).\label{eq:phiEl}
\end{align}

\subsection{Signal and baseline channel model} \label{sec:signalTransmissionStatic}
We consider the transmission of $L$ OFDM symbols with $N$ subcarriers. Under the assumption of perfect frequency synchronization between the UE and the BS, the received signal after the \ac{fft} operation at the UE in the frequency/slow-time domain can be represented by the matrix $\bm{Y} \in \mathbb{C}^{N\times L}$ as 
\begin{align}\label{eq:channelModel:WB}
    \bm{Y} &=  \bm{Y}_{\mathrm{b}}+ \bm{Y}_{\mathrm{r}}+\bm{N},
\end{align}
where the noise matrix is represented with $\bm{N}$, whose elements are drawn independently from a circularly symmetric Gaussian distribution with variance $N_0$. The matrices $\bm{Y}_{\mathrm{b}}$ and $\bm{Y}_{\mathrm{r}}$ describe the signal received through the direct and reflected path, respectively. 
 As a baseline, we consider a channel model that ignores any spatial-WB effect and assumes a sufficiently short observation time such that approximately $\bm{v}=\bm{0}$. For simplicity, we assume that all the transmitted symbols are equal to one. Hence following  \cite{keykhosravi2020siso}
 %Based on \eqref{eq:CNmatrix} and \eqref{eq:Ematrix}, we have that $[\bm{C}_{\mathrm{d}}(\bm{0})]_{n,\ell}=1$ for all $\ell$ and $n$, and $\bm{E}(\bm{0}) = \bm{I}_{N}$. Substituting these into \eqref{eq:Yb} and \eqref{eq:Yr}, the received signal is modeled as
 %\begin{align}
 %\bm{Y}_{\mathrm{s}}=\bm{Y}_{\mathrm{bs}}+\bm{Y}_{\mathrm{rs}},\label{eq:YStatic}
 %\end{align}
 \begin{align}
    \bm{Y}_{\mathrm{b}} &= g_{\mathrm{b}} \bm{D}(\tau_{\mathrm{b}})\label{eq:YbStatic} \\ 
    \bm{Y}_{\mathrm{r}} &=  g_{\mathrm{r}} \bm{D}(\tau_\mathrm{r}) \odot \bm{A}( \bm{\phi}),\label{eq:YrStatic} 
\end{align}
where the complex channel gain for the direct path is indicated by $g_{\mathrm{b}}$ and for the reflected one by $g_{\mathrm{r}}$. The matrix  $\bm{D} \in \complexset{N}{L}$ is the delay steering vector repeated across time and is defined as
\begin{align}\label{eq:matrixD}
    \bm{D}(\tau) = [1, e^{-\jmath 2\pi\deltaf \tau }, \dots, e^{-\jmath 2 \pi (N-1) \deltaf \tau }]^\top \bm{1}_{L}^{\top},
\end{align}
  where $\deltaf$ is the subcarrier spacing. Let $\thetab$ denote the known \ac{aoa} from the \ac{bs} to the \ac{ris}. In \eqref{eq:YrStatic}, $\bm{A}(\bm{\phi})\in \complexset{N}{L}$ captures the effects of \ac{ris} phase modulation, given by
\begin{align}
    [\bm{A}(\bm{\phi})]_{n,\ell} &= \bm{a}(\thetab)^{\top} \diag(\bm{\gamma}_{\ell}) \bm{a}(\phib),\label{eq:aphi}
\end{align}
where all the rows of $\bm{A}(\bm{\phi})$ are identical. The vector $\bm{\gamma}_{\ell} \in \mathrm{C}^{M}$ is defined as
\begin{align}
    \bm{\gamma}_{\ell} = \mathrm{vec}(\bm{\Gamma}_{\ell})\label{eq:def_gamma}
\end{align}
 and it represents the RIS phase profile vector at time $\ell$. The vector $\bm{a}(\cdot) \in \mathbb{C}^{M}$ is the narrowband \ac{ris} response steering vector  and is defined as
\begin{align} 
    [\bm{a}(\bm{\psi})]_{m} = \exp\left(\jmath\bm{k}(\bm{\psi})^\top [\bm{Q}]_{:,m}\right),\label{eq:aVector}
\end{align}
where the relative RIS element positions are contained in
\begin{align}\label{eq:Q}
    \bm{Q}&=[\bm{q}_{0,0},\bm{q}_{1,0}, \dots, \bm{q}_{M_1-1,M_2-1}].
\end{align}
The wavenumber vector is defined as
\begin{align}\label{eq:WaveNumVect}
    \bm{k}(\bm{\psi}) &= \frac{2\pi}{\lambda}[\sin([\bm{\psi}]_{\mathrm{el}})\cos([\bm{\psi}]_{\mathrm{az}}),\nonumber\\
    &\qquad\qquad \quad  \sin([\bm{\psi}]_{\mathrm{el}})\sin([\bm{\psi}]_{\mathrm{az}}),\cos([\bm{\psi}]_{\mathrm{el}})]^\top,
\end{align}
where $[\bm{\psi}]_{\mathrm{az}}$ and $[\bm{\psi}]_{\mathrm{el}}$ represent the azimuth and elevation of the generic direction described by angle $\bm{\psi}$ (see Fig.\,\ref{fig:setup}(b)), and $\lambda=c/f_c$ is the wavelength at the carrier frequency.

\section{Extended channel models for spatial-WB and UE mobility} \label{sec:extended-channel-models}

While the channel model from Section \ref{sec:signalTransmissionStatic} is common in the RIS literature, it is limited in two ways. First of all, when the RIS and the signal bandwidth are both large, the model fails to capture the variation of the RIS steering vector with the frequency, which is a consequence of the definition of the structure of $\bm{A}(\bm{\phi})$ with identical rows. Secondly, the assumption of negligible velocity severely limits the duration of the coherent processing interval $L/\deltaf$.

We now present two channel models that extend the model from Section \ref{sec:signalTransmissionStatic} in non-trivial ways: the first model captures both the spatial-WB effects \rev{\cite{dovelosintelligent,face_squint,chen2021beam}} and \ac{ue} mobility \rev{\cite{sun_wcl_doppler}}, and it is used for developing the \ac{crb} and simulating the channel; the second model neglects the spatial-WB effects and is employed in the estimator design. The original model \eqref{eq:YbStatic}--\eqref{eq:YrStatic}, which neglects both \ac{ue} mobility and spatial-WB effects, will be assumed for designing the \ac{ris} phase profiles.

\subsection{Signal transmission: Dynamic spatial-wideband model} \label{sec:signalTransmissionWB}
\rev{In the dynamic spatial-WB model, two fundamental changes occur with respect to the static spatial-NB model in \eqref{eq:YbStatic}--\eqref{eq:YrStatic}. First, the RIS response matrix $\bm{A}(\bm{\phi})$ in \eqref{eq:aphi} becomes frequency-dependent, leading to non-identical rows. Second, we incorporate new steering matrices that capture fast-time (sample-level) and slow-time (symbol-level) Doppler-induced phase progressions. Accordingly, as} shown in Appendix\,\ref{app:Specially_wideband_Ch_Model}, \eqref{eq:YbStatic}--\eqref{eq:YrStatic} should be extended to\footnote{We assume that the angular displacement caused by UE mobility is negligible due to the far-field assumption.}
\begin{align}
\bm{Y}_{\mathrm{b}} &= g_{\mathrm{b}} \bm{F}\bm{E}(\vub) \bm{F}^{\herm} \left(\bm{D}(\tau_{\mathrm{b}}) \odot \ccbigw(\vub)\right),  \label{eq:Yb}\\ 
\bm{Y}_{\mathrm{r}} &=  g_{\mathrm{r}} \bm{F} \bm{E}(\vurk) \bm{F}^{\herm} \left[\bm{D}(\tau_\mathrm{r}) \odot \aabigw( \bm{\phi}) \odot \ccbigw(\vurk)  \right].\label{eq:Yr}
\end{align}
Here, the matrix $\bm{F} \in \complexset{N}{N}$ is the unitary DFT matrix with elements 
 \begin{align}
     \left[ \bm{F} \right]_{n,\ell} = \frac{1}{\sqrt{N}} e^{- \jmath 2 \pi  \frac{n \ell}{N}}\label{eq:dft}
 \end{align} 
 for $n,\ell\in\{0,\dots,N-1\}$. In addition, $ \aabigw(\phib)$ represents the spatial-wideband version of $\aabig(\phib)$ in \eqref{eq:aphi}; namely,
 \begin{align}
    [\aabigw(\bm{\phi})]_{n,\ell} &= \bm{a}_n(\thetab)^{\top} \diag(\bm{\gamma}_{\ell}) \bm{a}_n(\phib),\label{eq:aphi_wb}
\end{align}
where the RIS steering vector now depends on the subcarrier index $n$:
\begin{align} 
    [\bm{a}_n(\bm{\psi})]_{m} = \exp\left(\jmath\bm{k}_n(\bm{\psi})^\top [\bm{Q}]_{:,m}\right),\label{eq:aVector_wb}
\end{align}
with $\bm{k}_n(\bm{\psi})$ being defined as in \eqref{eq:WaveNumVect} by replacing $\lambda$ with 
%\begin{align}
 %   [\bm{A}(\bm{\phi})]_{n,\ell} &= \bm{a}_{n}(\theta)^{\top} \diag(\bm{\gamma}_{\ell}) \bm{a}_{n}(\phi),\label{eq:aphi}
%\end{align}
%where $\theta$ is the known \ac{aoa} from \ac{bs} to the \ac{ris}. The vector $\bm{\gamma}_{\ell} \in \mathrm{C}^{M}$ is defined as
%\begin{align}
 %   \bm{\gamma}_{\ell} = \mathrm{vec}(\bm{\Gamma}_{\ell})\label{eq:def_gamma}
%\end{align}
 %and it represents the RIS phase profile vector at time $\ell$. The vector $\bm{a}_{n}(\cdot) \in \mathbb{C}^{M}$ is the \ac{ris} response steering vector for the $n$th subcarriers and is defined as
%\begin{align} 
 %   [\bm{a}_n(\bm{\psi})]_{m} = \exp\left(\jmath\bm{k}_n(\bm{\psi})^\top [\bm{Q}]_{:,m}\right),\label{eq:aVector}
%\end{align}
%where 
%%\begin{align}\label{eq:Q}
 %   \bm{Q}&=[\bm{q}_{0,0},\bm{q}_{1,0}, \dots, \bm{q}_{M_1-1,M_2-1}].
%\end{align}
%The wavenumber vector is defined as
%\begin{align}\label{eq:WaveNumVect}
 %   \bm{k}_n(\bm{\psi}) &= \frac{2\pi}{\lambda_n}[\sin([\bm{\psi}]_{\mathrm{el}})\cos([\bm{\psi}]_{\mathrm{az}}),\nonumber\\
 %   &\qquad\qquad \quad  \sin([\bm{\psi}]_{\mathrm{el}})\sin([\bm{\psi}]_{\mathrm{az}}),\cos([\bm{\psi}]_{\mathrm{el}})]^\top,
%%\end{align}
%where $[\bm{\psi}]_{\mathrm{az}}$ and $[\bm{\psi}]_{\mathrm{el}}$ represent the azimuth and elevation of the generic angle $\bm{\psi}$ (see Fig.\,\ref{fig:setup}(b)).
\begin{align}
    % \lambda_n = \frac{c}{f_c+\left(n-(N-1)/2\right)\deltaf}.\label{eq:lambda_n}
        \lambda_n = \frac{c}{f_c+n\deltaf}.\label{eq:lambda_n}
\end{align}
\rev{Moreover}, the effects of \ac{ue} mobility on the received signal is captured by the \ac{ici} phase rotation matrix $\bm{E}(v) \in \complexset{N}{N}$, \rev{which models Doppler-induced \textit{fast-time} phase rotations within an OFDM symbol \cite{Visa_CFO_TSP_2006,multiCFO_TSP_2019,multiCFO_TCOM_2019}}, and the temporal steering matrix $\ccbigw(v)\in \complexset{N}{L}$, \rev{which quantifies Doppler-induced \textit{slow-time} phase progressions across consecutive OFDM symbols \cite{OFDM_ICI_TVT_2017,ICI_Friend_Foe_JSTSP}}:
\begin{align}  
	[\ccbigw(v)]_{n,\ell} & \triangleq   e^{\jmath 2\pi \ell \Tsym  v/\lambda_n } \label{eq:Cmatrix}
	\\ 
	\bm{E}(v) &\triangleq \diagg{1, e^{\jmath 2 \pi \frac{\Tso}{N}  v/\lambda }, \ldots, e^{\jmath 2 \pi  \frac{\Tso(N-1)}{N}  v/\lambda} } \label{eq:Ematrix}
\end{align}
 for $n\in\{0,\dots,N-1\}$ and $\ell\in\{0,\dots,L-1\}$. Here, $\Tso=1/\deltaf$ is the elementary symbol duration and $\Tsym = \Tcp + \Tso$ is the total signal duration, with $\Tcp$ denoting the cyclic prefix (CP) duration.

 %%%%%%%%%%%%%%%%%%%%%%%%%%%%%%%%%%%%%%%%
 
 \subsection{Signal transmission: Dynamic spatial-narrowband model} \label{sec:signalTransmissionNB}
 In order to reduce the complexity of our estimator, we design it based on a simpler channel than \eqref{eq:Yb}--\eqref{eq:Yr} \rev{by assuming a spatial-narrowband model}. \rev{In this case,} the channel \rev{in \eqref{eq:Yb}--\eqref{eq:Yr}} is constructed by reverting $\lambda_n$ in \eqref{eq:lambda_n} back to $\lambda=c/f_c$. This will simplify the structure of matrices $\ccbigw$ and $\aabigw$ by making their elements independent of \rev{the subcarrier index} $n$, i.e., all of their rows become identical. Specifically, \rev{under the spatial-narrowband model,} the received signal \rev{in \eqref{eq:Yb}--\eqref{eq:Yr} specializes to} 
\begin{align}
\bm{Y}_{\mathrm{b}} &= g_{\mathrm{b}} \bm{F}\bm{E}(\vub) \bm{F}^{\herm} \left(\bm{D}(\tau_{\mathrm{b}}) \odot \ccbig(\vub)\right)  \label{eq:Ybn}\\ 
\bm{Y}_{\mathrm{r}} &=  g_{\mathrm{r}} \bm{F} \bm{E}(\vurk) \bm{F}^{\herm} \left[\bm{D}(\tau_\mathrm{r}) \odot \aabig( \bm{\phi}) \odot \ccbig(\vurk) \right] ~,\label{eq:Yrn}
\end{align}
\rev{where the subcarrier-dependent matrices $\ccbigw(v)$ and $\aabigw(\bm{\phi})$ in \eqref{eq:Yb}--\eqref{eq:Yr} revert to their narrowband (subcarrier-independent) counterparts $\ccbig(v)$ and $\aabig(\bm{\phi})$. Here,} $\aabig(\bm{\phi})$ is defined in \eqref{eq:aphi} and\rev{\footnote{\rev{Note that the dynamic spatial-narrowband model \eqref{eq:Ybn}--\eqref{eq:Yrn} reverts to the static spatial-narrowband model \eqref{eq:YbStatic}--\eqref{eq:YrStatic} when $v = 0$.  }}} 
%Here, we explicitly introduce the subscript $\mathrm{d}$ to distinguish from the more general model \eqref{eq:Yb}--\eqref{eq:Yr}:
\begin{align}
	[\ccbig(v)]_{n,\ell} & \triangleq   e^{\jmath 2 \pi \ell \Tsym  v/\lambda } \label{eq:CNmatrix}.
\end{align}
%where 
%\begin{align}
 %   \bm{a}(\bm{\psi}) &= \exp\left(\jmath\bm{k}(\bm{\psi})^\top \bm{Q}\right)\label{eq:aVectorNB}\\
  %  \bm{k}(\bm{\psi}) &= \frac{2\pi}{\lambda}[\sin([\bm{\psi}]_{\mathrm{el}})\cos([\bm{\psi}]_{\mathrm{az}}),\nonumber\\
   % &\qquad\qquad \quad  \sin([\bm{\psi}]_{\mathrm{el}})\sin([\bm{\psi}]_{\mathrm{az}}),\cos([\bm{\psi}]_{\mathrm{el}})]^\top.\label{eq:WaveNumVect2}
%\end{align}

\rev{For the spatial-narrowband approximation in \eqref{eq:Ybn} and \eqref{eq:Yrn} to be valid, the following conditions must be satisfied (see Appendix~\ref{app_nb_valid} for details):}
%To derive \eqref{eq:Ybn} and \eqref{eq:Yrn} from the more general model \eqref{eq:Yb} and \eqref{eq:Yr}, we use the following assumptions, respectively:
\begin{align}
    \max\{v_{\mathrm{r}},v_{\mathrm{b}}\} L\Tsym B \approx \max\{v_{\mathrm{r}},v_{\mathrm{b}}\}LN&\ll c 
    \label{eq:condVelWB}
    \\
    \max(M_1,M_2)d \sin(\alpha) B &\ll c ~, \label{eq:condSpWB}
\end{align}
\rev{which ensure the validity of the approximations $\ccbigw(v) \approx \ccbig(v)$ and $\aabigw(\bm{\phi}) \approx \aabig(\bm{\phi})$, respectively.}
  Here, $\alpha=\max\{\alpha_{\mathrm{\phi}},\alpha_{\mathrm{\theta}}\}$, where  $\alpha_{\mathrm{\phi}}$ and $\alpha_{\mathrm{\theta}}$ are the angles between the RIS normal ($[0,1,0]^\top$) and the two vectors $\bm{k}(\bm{\phi})$ and $\bm{k}(\bm{\theta})$, respectively, which are defined in \eqref{eq:WaveNumVect}.
While the condition in \eqref{eq:condVelWB} almost always holds (corresponding to the assumption of small time-bandwidth product \cite{OFDM_ICI_TVT_2017}), the condition in \eqref{eq:condSpWB} does not hold for RISs with large dimension combined with signals of large bandwidth \cite{wang2018spatial}. We will study the effects of this assumption in Section~\ref{sec:simulationResults}.

\section{RIS phase profile design}\label{sec:RisPhaseDesign}
In this section, we consider the design of the \ac{ris} phase profile $\bm{\Gamma}_{\ell}$ for $\ell = 0, \dots L-1$. In order to mitigate the interference between the direct path and the reflected one, we use the method described in \cite{keykhosravi2021multi}. The method deploys temporal orthogonal \ac{ris} phase profiles and a post processing at the receiver. \rev{This process resembles the code-division multiplexing, which is a well-known method in wireless communications (see e.g., \cite{hwacdma}).} It can remove the interpath interference completely in the static scenario ($\bm{v}=\bm{0}$).
%However, in this paper we use this method in the dynamic scenario to mitigate the interference partially. The residual interference is then compensated for via the estimator using successive cancellation.
Next, we use the static channel model in Section\,\ref{sec:signalTransmissionStatic} to describe the \ac{ris} phase profile design.

\subsection{Orthogonal RIS phase profiles}
We set $L$ to be an even number and  for each $ k = 0, 1, \dots, L/2$  we select beams $\bm{B}_{k}\in\mathbb{C}^{M_1\times M_2}$  either randomly or according to a directional codebook (we elaborate on this in Section\,\ref{sec:beamforming}). Also, similarly as in \eqref{eq:def_gamma}, we define $\bm{b}_k = \mathrm{vec}(\bm{B}_k)$. Then we set $\bm{\gamma}_{2k} = \bm{b}_{k}$ and  $\bm{\gamma}_{2k+1} = - \bm{b}_{k}$. By doing so, from \eqref{eq:aphi} we have that $[\bm{A}(\bm{\phi})]_{:,2k+1} = -[\bm{A}(\bm{\phi})]_{:,2k}$. Therefore, from \eqref{eq:YbStatic} and \eqref{eq:YrStatic}, we have
 \begin{align}
    [\bm{Y}_{\mathrm{b}}]_{:,2k+1} & =g_b \left[\bm{D}(\tau_{\mathrm{b}})\right]_{:,2k+1}\\
    &= [\bm{Y}_{\mathrm{b}}]_{:,2k} \label{eq:Yb_2l}\\
    [\bm{Y}_{\mathrm{r}}]_{:,2k+1} & = g_{\mathrm{r}} \left[\bm{D}(\tau_\mathrm{r})\right]_{:,2k+1} \odot \left[\bm{A}( \bm{\phi})\right]_{:,2k+1}  \\
    & = -  g_{\mathrm{r}} \left[\bm{D}(\tau_\mathrm{r})\right]_{:,2k} \odot \left[\bm{A}( \bm{\phi})\right]_{:,2k}\label{eq:Yr2l0}\\
    &=-[\bm{Y}_{\mathrm{r}}]_{:,2k}.\label{eq:Yr2l}
\end{align}
The post-processing step at the receiver involves calculating matrices $\bm{Z}_{\mathrm{b}}\in\complexset{N}{L/2}$ and $\bm{Z}_{\mathrm{r}}\in\complexset{N}{L/2}$ as 
\begin{align}
    \left[\bm{Z}_{\mathrm{b}}\right]_{:,k} &= \left[\bm{Y}\right]_{:,2k} + \left[\bm{Y}\right]_{:,2k+1}\\
    & = 2g_{\mathrm{b}} \left[\bm{D}(\tau_{\mathrm{b}})\right]_{:,2k}+\left[\bm{N}\right]_{:,2k}+\left[\bm{N}\right]_{:,2k+1}\\
    &= 2[\bm{Y}_{\mathrm{b}}]_{:,2k}+\left[\bm{N}\right]_{:,2k}+\left[\bm{N}\right]_{:,2k+1}\label{eq:Zbs}\\
    \left[\bm{Z}_{\mathrm{r}}\right]_{:,k} &= \left[\bm{Y}\right]_{:,2k} - \left[\bm{Y}\right]_{:,2k+1}\\
    & = 2g_{\mathrm{r}} \left[\bm{D}(\tau_\mathrm{r})\right]_{:,2k} \odot \left[\bm{A}( \bm{\phi})\right]_{:,2k}+\left[\bm{N}\right]_{:,2k}-\left[\bm{N}\right]_{:,2k+1}\nonumber\\
    &= 2[\bm{Y}_{\mathrm{r}}]_{:,2k}+\left[\bm{N}\right]_{:,2k}-\left[\bm{N}\right]_{:,2k+1}.\label{eq:Zrs}
\end{align}
It can be seen from \eqref{eq:Zbs} and \eqref{eq:Zrs} that the matrix $\bm{Y}_{\mathrm{b}}$  ($\bm{Y}_{\mathrm{r}}$)  depends only on the parameters of the direct (reflected) channel. Therefore, with the aforementioned \ac{ris} phase profile design and post-processing, we can remove the interference between the two paths, which facilitates the estimation of the channel parameters. \rev{Furthermore, from \eqref{eq:Zbs} and \eqref{eq:Zrs} it can be seen the signals $\bm{Z}_{\mathrm{b}}$  and $\bm{Z}_{\mathrm{r}}$ have higher SNRs compared to the signals $\bm{Y}_{\mathrm{b}}$  and $\bm{Y}_{\mathrm{r}}$, respectively. This indicates that the presented orthogonal coding does not result in a waste of resources by repeating the beams.} For clarification, we consider a toy example with $L=4$ and $M=1$ and  $[\bm{b}_0, \bm{b}_1]=[e^{\jmath \theta_0},e^{\jmath \theta_1}]$ for some $\theta_0,\theta_1 \in [0\,2\pi)$. Then the set of RIS phase profiles would be $[\bm{\gamma}_0, \bm{\gamma}_1, \bm{\gamma}_2, \bm{\gamma}_3] = [e^{\jmath \theta_0}, - e^{\jmath \theta_0}, e^{\jmath \theta_1}, -e^{\jmath \theta_1}]$. Also if the noise is neglected, we have $[\bm{Z}_{\mathrm{b}}]_{:,k} = 2 g_{\mathrm{b}} \bm{d}(\tau_{\mathrm{b}})$ and $[\bm{Z}_{\mathrm{r}}]_{:,k} = 2 g_{\mathrm{r}} e^{\jmath \theta_k}\bm{d}(\tau_{\mathrm{r}})$ for $k=0, 1$. 

For future use, we refer to the post processing step in \eqref{eq:Zbs} and \eqref{eq:Zrs} as matching the signal $\bm{Y}$ with vectors $\bm{w}_{\mathrm{b}}=[1,1]^{\top}$ and $\bm{w}_{\mathrm{r}}=[1,-1]^{\top}$, respectively. We explain this step in Algorithm\,\ref{alg:match} as follows.
%------------------------------match algorithm---------------
\begin{algorithm}[h]
	\caption{\textit{match($\bm{Y}$,$\bm{w}$)} }\label{alg:match}
	\textbf{Inputs:} Received signal ($\bm{Y}\in \complexset{N}{L}$) and vector $\bm{w}\in \mathbb{C}^2$. \\
	\textbf{Output:} $\bm{Z}\in \complexset{N}{L/2}$.
	\begin{algorithmic}[1]
		\For {$k\in \{0, \dots, L/2-1\}$}
		\State $\left[\bm{Z}\right]_{:,k} = [\bm{w}]_1\left[\bm{Y}\right]_{:,2k} + [\bm{w}]_2 \left[\bm{Y}\right]_{:,2k+1}$
		\EndFor
		\Return $\bm{Z}$
	\end{algorithmic}
\end{algorithm}
%------------------------------end of match algorithm---------------
\subsection{Loss of orthogonality due to UE mobility}

For the dynamic case ($\bm{v}\neq \bm{0}$), one can write \rev{\eqref{eq:Zbs}--\eqref{eq:Zrs}} as
\begin{align}
    \left[\bm{Z}_{\mathrm{b}}\right]_{:,k} &=(2-\epsilon(v_{\mathrm{b}}))[\bm{Y}_{\mathrm{b}}]_{:,2k}+\epsilon(v_{\mathrm{r}})[\bm{Y}_{\mathrm{r}}]_{:,2k}\nonumber\\
    & \ \ \ \ +\left[\bm{N}\right]_{:,2k}+\left[\bm{N}\right]_{:,2k+1}\label{eq:Zbs_dynamic}\\
    \left[\bm{Z}_{\mathrm{r}}\right]_{:,k} &= (2-\epsilon(v_{\mathrm{r}}))[\bm{Y}_{\mathrm{r}}]_{:,2k}+\epsilon(v_{\mathrm{b}})[\bm{Y}_{\mathrm{b}}]_{:,2k}\nonumber\\
    & \ \ \ \ +\left[\bm{N}\right]_{:,2k}-\left[\bm{N}\right]_{:,2k+1}\label{eq:Zrs_dynamic},
\end{align}
where $\epsilon(v) = 1-\exp(\jmath 2\pi T_{\mathrm{sym}} v/\lambda)$. By comparing \eqref{eq:Zbs_dynamic}--\eqref{eq:Zrs_dynamic} with \eqref{eq:Zbs}--\eqref{eq:Zrs}, one can see that \ac{ue} mobility introduces two impairments to the proposed method:
\begin{itemize}
    \item Energy loss: some of the signal energy of the desired path is lost since $\vert2-\epsilon(v)\vert<2$
    \item Residual interference: the second term in \eqref{eq:Zbs_dynamic}--\eqref{eq:Zrs_dynamic} exhibits the interference from the undesired path.
\end{itemize}
We design our estimator based on the approximation $\epsilon(v) = 0$. To counter the aforementioned impairments,  we apply multiple iterations and deploy successive cancellation.

\subsection{\rev{RIS phase profile design}}\label{sec:beamforming}
In this section, we discuss the selection of $\bm{B}_{k}$ or equivalently $\bm{b}_{k}$. We consider two methods, namely random and directional profiles. The latter can be used when a prior information about the \ac{ue} location is available and the former when such information is lacking. 

\subsubsection{Random profile}\label{sec:randCodebook} With the random codebook, for $m = 0,\dots,M-1$ and $k = 0,\dots,L/2-1$  we let
\begin{align}
    [\bm{b}_{k}]_m = e^{\jmath \theta_{k,m}},
\end{align}
where $\theta_{k,m}$ are \ac{iid} realizations of the uniform distribution over the interval $[0,2\pi)$.

\subsubsection{Directional profile} \label{sec:dirCodebook}
Here, we assume that we have a prior knowledge of the \ac{ue} position, $\bm{\xi}$, which  is distributed uniformly throughout the sphere
\begin{align}
  \vert \bm{p}-\bm{\xi}\vert<\sigma.\label{eq:sphere}
\end{align}
  We call $\sigma$ the uncertainty radius. Given the prior position knowledge $\bm{\xi}$, the \ac{ris} phase profile is designed as follows. We first select $L/2$ points $\bm{\xi}_{0},\dots,\bm{\xi}_{L/2-1}$ randomly (with uniform distribution) from the sphere centered at  $\bm{\xi}$ with radius $\sigma$. Second, we set $\bm{b}_{k} = \bm{f}(\bm{\xi}_{k})$ where for $m=0,\dots, M-1$:
 \begin{align}
     [\bm{f}(\bm{x})]_{m} = \exp\left(-\jmath \left(\bm{k}(\bm{\theta})^{\top} +\frac{2\pi(\bm{x}^{\top}-\bm{p}_{\mathrm{r}}^{\top})}{\lambda\Vert \bm{x}-\bm{p}_{\mathrm{r}} \Vert}\right) [\bm{Q}]_m\right). \label{eq:GammaFunction}
 \end{align}
One can see that with the phase profile in \eqref{eq:GammaFunction}  the reflected signal energy from the \ac{ris} is concentrated towards the point $\bm{x}$.

\section{Estimation algorithm}\label{sec_estimator}
\begin{figure}
    \centering
    \includegraphics[width=\columnwidth]{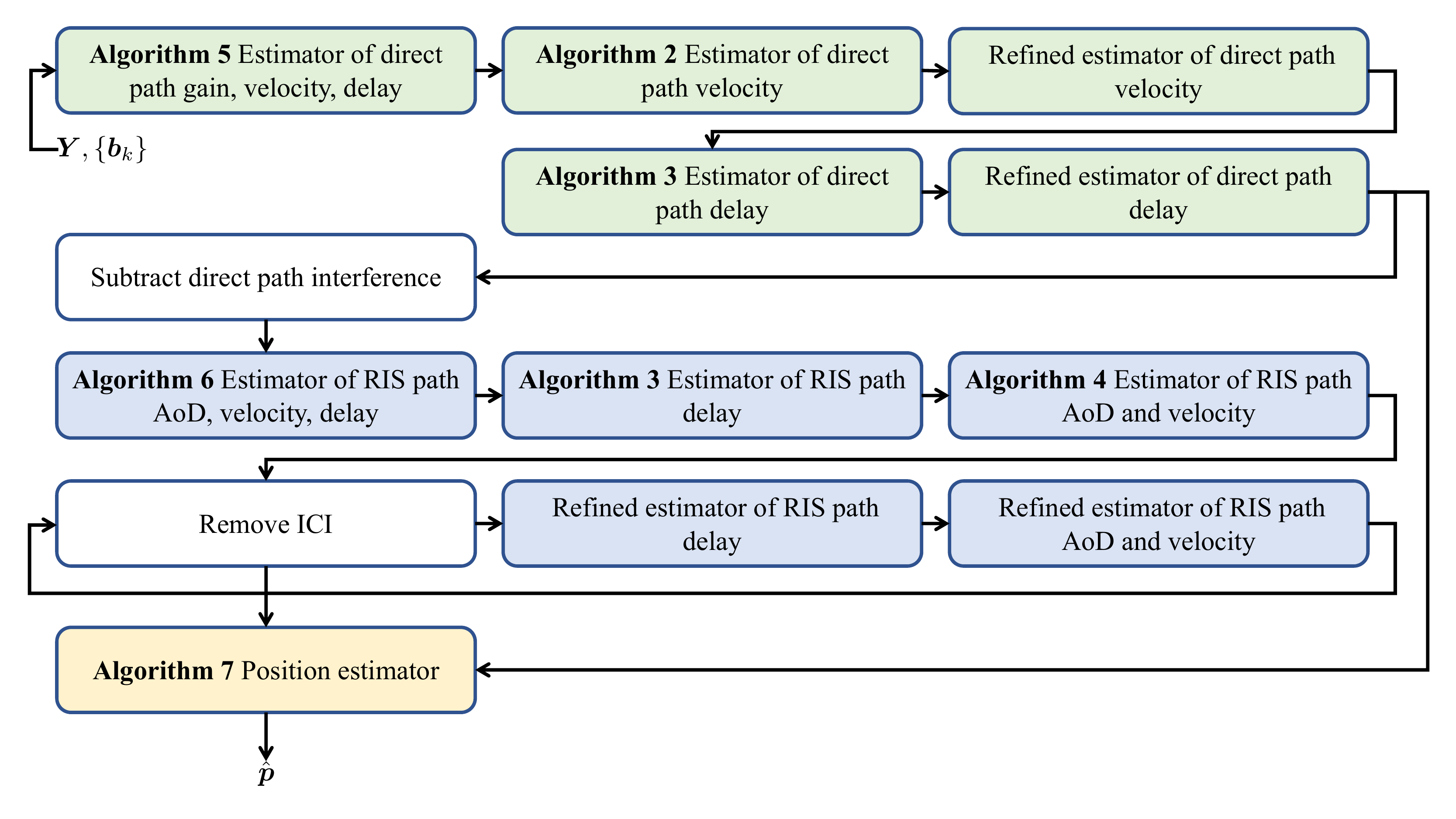}
    \caption{A flowchart of Algorithm~\ref{alg:estimator}, comprising three stages: estimation of the parameters of the direct path (green), estimation of the parameters of the reflected path (blue), and position estimation (orange).}
    \label{fig:flowchart}
\end{figure}

In this section, we propose an estimator to estimate first the channel parameters and then the \ac{ue} position and clock bias based on them. The overall  process is described in the flowchart of Fig.\,\ref{fig:flowchart} using multiple separate procedures described in Algorithm~\ref{alg:coarse_v}--\ref{alg:pos} as building  blocks. \rev{ To estimate the parameters, we first obtain a coarse estimation and then use it as an initial point in a refinement process, which is a standard approach in localization literature (see e.g., \cite{Shahmansoori18TWC}). Next we describe the Algorithms using a bottom-up approach.}

\subsection{Estimation of $v_{\mathrm{b}}$} \label{sec:estimation_v}
For estimating the UE velocity, we first obtain a coarse estimation using standard methods based on the \ac{dft} matrix and then provide a refined estimation by using the coarse estimate as the initial point for our  optimization.
%For the ease of presentation, we first describe the latter step and then the former one.

 \rev{Algorithm\,\ref{alg:coarse_v} provides a coarse  estimation of the velocity.} The input signal is an estimate of $\bm{Z}_{\mathrm{b}} = \textit{match}(\bm{Y},\bm{w}_{\mathrm{b}})$ described in \eqref{eq:Zbs_dynamic}. One can see that for every $n$ we have that
\begin{align}
    [\bm{Z}_{\mathrm{b}}]_{n,:} \approx \xi_n [ 1, e^{\jmath 2h_{\mathrm{v}}v_{\mathrm{b}}}, e^{\jmath 4h_{\mathrm{v}}v_{\mathrm{b}}}, \dots, e^{\jmath (L-2) h_{\mathrm{v}}v_{\mathrm{b}}} ]\label{eq:Zb_n}
\end{align}
for some scalar $\xi_n\in \mathbb{C}$, \rev{ where $h_{\mathrm{v}} = 2\pi T_{\mathrm{sym}}/\lambda$.} Then it can be seen that the maximum of \rev{
\begin{align}\label{eq:fv_defenition}
    f(v) = \Vert \hat{\bm{Z}}_{\mathrm{b}}[ 1, e^{\jmath 2h_{\mathrm{v}}v}, \dots, e^{\jmath (L-2) h_{\mathrm{v}}v} ]^{\herm} \Vert^2
\end{align}}
provides an estimate of $v_{\mathrm{b}}$. \rev{To find the maximum of $f(v)$, we note that} the structure shown in \eqref{eq:Zb_n} is similar to the rows of the \ac{dft} matrix $\bm{F}$ in Line\,\ref{CoarseV:Line1} of Algorithm\,\ref{alg:coarse_v}, that is
\begin{align}
    [\bm{F}]_{n,:}=\frac{\sqrt{2}}{\sqrt{L}}[1, e^{-\jmath \omega n}, e^{-\jmath 2\omega n}, \dots, e^{-\jmath (L/2-1)\omega n}]\label{eq:FRowsForV}
\end{align}
for all $n= 0, \dots, N_{\mathrm{v}}-1$. Here, $\omega=2\pi/N_{\mathrm{v}}$ and $N_{\mathrm{v}}$ is a design parameter that determines the dimension of the \ac{dft} matrix and accuracy of our coarse estimation. By comparing \eqref{eq:FRowsForV} to \eqref{eq:Zb_n}, one can \rev{approximate the $\arg\max_v f(v)$} via the maximization in Line\,\ref{CoarseV:Line3} and the assignment in Line\,\ref{CoarseV:Line6}. Finally, the condition in Line\,\ref{CoarseV:Line4} compensates for the wrap-around effect in the complex-exponential function when $v_{\mathrm{b}}<0$.

\rev{\emph{Refinement}: Let the output of the Algorithm\,\ref{alg:coarse_v} be $v_{0}$. To refine this estimation, we perform the maximization $\hat{v}_b=\max_v f(v)$ via a \emph{quasi-Newton} algorithm initiated at $v_{0}$.}
\begin{algorithm}[h]
	\caption{\textit{Coarse\_Velocity\_Est($\hat{\bm{Z}}_{\mathrm{b}}$)} }\label{alg:coarse_v}
	\textbf{Inputs:}  Signal ($\hat{\bm{Z}}_{\mathrm{b}}\in \complexset{N}{L/2}$) \\
	\textbf{Parameters:}  DFT dimension ($N_{\mathrm{v}}$) \\
	\textbf{Output:} $\hat{v}_{0}$
	\begin{algorithmic}[1]
		\State $ \bm{F} \gets N_{\mathrm{v}} \times L/2$ DFT matrix\label{CoarseV:Line1}
		\State $\bm{Z}_{\mathrm{v}} \gets \bm{F} \hat{\bm{Z}}_{\mathrm{b}}^{\top}$\label{CoarseV:Line2}
		\State $i_{\mathrm{m}} \gets \mathrm{argmax}_i \Vert [\bm{Z}_{\mathrm{v}}]_{i,:}\Vert$ \label{CoarseV:Line3}
		\If{$i_{\mathrm{m}}> N_{\mathrm{v}}/2$}\label{CoarseV:Line4}
		\State $i_{\mathrm{m}} \gets i_{\mathrm{m}} - N_{\mathrm{v}}+1$\label{CoarseV:Line5}
		\EndIf
		\State $\hat{v}_{0} \gets i_{\mathrm{m}}\lambda/(2T_{\mathrm{sym}}N_{\mathrm{v}})$\label{CoarseV:Line6}\\
		\Return $\hat{v}_{0}$
	\end{algorithmic}
\end{algorithm}
%------------------------------end of Coarse velocity Est algorithm---------------

\subsection{Estimation of ToA}\label{sec:Estimation_tau}
Similar to Section\,\ref{sec:estimation_v}, the estimation of \ac{toa} comprises  coarse and fine estimation steps. 

\rev{Algorithm\,\ref{alg:coarse_tau} describes the coarse} estimation of the \ac{toa} given the input signal $\bm{Z}_{\mathrm{\tau}}$\footnote{\rev{We use Algorithm\,\ref{alg:coarse_tau} within  Algorithm\,\ref{alg:estimator_direct}, where the input ($\bm{Z}_{\mathrm{\tau}}$) is an estimate of $\sum_t[\bm{Z}_{\mathrm{b}}]_{:,t}$ with dimension $N\times 1$  and also in Algorithm\,\ref{alg:estimator_reflected}, where the input is an estimate of $\bm{Z}_{\mathrm{r}}$ with dimension $N\times L/2$.}}. We assume that the columns of the input signal have the structure 
\begin{align}
    [\bm{Z}_{\mathrm{\tau}}]_{:,t}\approx \xi_{t}[1, e^{\jmath h_{\mathrm{\tau}}\tau_{\mathrm{x}}}, \dots, e^{\jmath (N-1) h_{\mathrm{\tau}}\tau_{\mathrm{x}}}]^\herm \label{eq:Y_tau_t}
\end{align}
for some $\xi_t\in\mathbb{C}$, where  $\tau_{\mathrm{x}}$ represents either $\{\tau_{\mathrm{b}}$ or $\tau_{\mathrm{r}}\}$.
 Algorithm\,\ref{alg:coarse_tau}  can be explained similarly as in Section\,\ref{sec:estimation_v} using \eqref{eq:Y_tau_t}.

\rev{\emph{Refinement:}}
 Based on \eqref{eq:Y_tau_t} a fine estimation of $\tau_{\mathrm{b}}$ or $\tau_{\mathrm{r}}$ can be found by calculating 
 \begin{align}\label{eq:fineTau}
     \hat{\tau}=\arg\max_{\tau} \Vert [ 1, e^{\jmath h_{\mathrm{\tau}}\tau}, \dots, e^{\jmath (N-1) h_{\mathrm{\tau}}\tau} ] \bm{Z}_{\mathrm{\tau}} \Vert^2,
 \end{align}
 where $h_{\mathrm{\tau}} \gets 2\pi \deltaf$. \rev{The optimization \eqref{eq:fineTau} can be solved via a quasi-Newton algorithm that uses the coarse estimation as the initial point of search.}

%------------------------------ Coarse delay Est algorithm---------------
\begin{algorithm}[h]
	\caption{\textit{Coarse\_delay\_Est($\bm{Z}_{\mathrm{\tau}}$)} }\label{alg:coarse_tau}
	\textbf{Inputs:}  Signal ($\bm{Z}_{\mathrm{\tau}}\in \complexset{N}{T}$, \rev{where $T\in \{1, L/2\}$})\\
	\textbf{Parameters:}  IDFT dimension ($N_{\tau}$) \\
	\textbf{Output:} $\hat{\tau}$
	\begin{algorithmic}[1]	
	\State $ \bm{F} \gets  N_{\tau} \times N$ DFT matrix
	\State $\bm{W}_{\tau} \gets \bm{F}^{\herm} \bm{Z}_{\mathrm{\tau}}$
	\State $i_{\mathrm{m}} \gets \mathrm{argmax}_i \Vert [\bm{W}_{\tau}]_{i,:}\Vert$
	\State $\hat{\tau}\gets i_{\mathrm{m}}/(\deltaf N_{\tau})$
	\\
	\Return $\hat{\tau}$
	\end{algorithmic}
\end{algorithm}
%------------------------------end of Fine velocity Est algorithm---------------

\subsection{Joint estimation of velocity and AoD for the reflected path}\label{sec:Estimation_va}
In this section, we describe coarse and fine steps for joint estimation of the angle and velocity. 

Algorithm\,\ref{alg:coarse_vA_phi} describes the \rev{coarse} estimation process of \ac{aod} and  velocity. We assume that the input  signal $\bm{z}_{\mathrm{\phi}}$ is proportional to the rows of the matrix $\bm{C}(v)\odot\bm{A}(\bm{\phi})$ and therefore has the structure 
\begin{align}
    [\bm{z}_{\mathrm{\phi}}]_k & = \xi e^{\jmath 2kh_{\mathrm{v}} v} \bm{a}(\bm{\theta})^\top \diag(\bm{b}_{k})\bm{a}(\bm{\phi})\label{eq:zPhi2}
\end{align}
for some constant $\xi\in\mathbb{C}$ and  velocity $v$, which is to be estimated. Also, the constant $h_{\mathrm{v}}$ is defined as $h_{\mathrm{v}} = 2\pi T_{\mathrm{sym}}/\lambda$. 

To obtain a coarse estimation of $v$ and $\bm{\phi}$ based on the input signal $\bm{z}_{\mathrm{\phi}}$ described in \eqref{eq:zPhi2}, Algorithm\,\ref{alg:coarse_vA_phi} uses a set of candidate \acp{aod}. For the $s$th candidate, we calculate $\bm{z}_{s}$ in Line\,\ref{CoarseVA_Line4} and then normalize it in Line\,\ref{CoarseVA_Line5} to obtain $\bm{w}_{s}$. Assume that for some $s_{\mathrm{m}}$ we have $\bm{\phi}_{s_\mathrm{m}} = \bm{\phi}$, then we have that
\begin{align}
    \bm{z}_{\bm{\phi}} \propto [1, e^{\jmath  2 h_{\mathrm{v}} v}, \dots, e^{\jmath  (L-2) h_{\mathrm{v}} v}]^\top \odot \bm{w}_{s_{\mathrm{m}}}. \label{eq:Zs}
\end{align}
Motivated by the structure in \eqref{eq:Zs}, we compute the correlation of $\bm{z}_{\bm{\phi}}$ with all $\bm{w}_{s}$ and all of the rows of the \ac{dft} matrix in  Lines\,\ref{CoarseVA_Line6}--\ref{CoarseVA_Line7}. Then, we search over different values of $s$  and $i$ (which indicates the rows of the \ac{dft} matrix) to find the one with the highest correlation in Line\,\ref{CoarseVA_Line8}. We estimate $v_{\mathrm{r}}$ through Lines\,\ref{CoarseVA_Line10}--\ref{CoarseVA_Line12}, which are the same steps as in Lines\,\ref{CoarseV:Line4}--\ref{CoarseV:Line6} of Algorithm\,\ref{alg:coarse_v}. We explain in Appendix\,\ref{app:fft} how to choose the candidate AoDs.

\rev{\emph{Refinement:}}  
According to the \rev{\ac{rhs}} of \eqref{eq:zPhi2}, for $k\in\{0, \dots, L/2\}$, we define 
\begin{align}
[\bm{g}(v,\bm{\phi})]_{k} = e^{\jmath 2 k h_{\mathrm{v}} v} \bm{a}(\bm{\theta})^{\top} \diag(\bm{b}_{k}) \bm{a}(\bm{\phi}),
\end{align}
\rev{which is a function of  $v$ and $\bm{\phi}$}. 
 Then one can estimate the constant $\xi$ as
\begin{align}
    \hat\xi = {\bm{g}(v,\bm{\phi})^{\herm}\bm{z}_{\mathrm{\phi}}}/{\bm{g}(v,\bm{\phi})^{\herm}\bm{g}(v,\bm{\phi})}.
\end{align}
Next, we can define the objective function
\begin{align}
    f(v,\bm{\phi}) = \Vert\bm{z}_{\mathrm{\phi}} - \left({\bm{g}(v,\bm{\phi})^{\herm}\bm{z}_{\mathrm{\phi}}}/{\bm{g}(v,\bm{\phi})^{\herm}\bm{g}(v,\bm{\phi})}\right)\bm{g}(v,\bm{\phi})\Vert.
\end{align}
 \rev{To refine the estimation of $v$ and $\bm{\phi}$, we conduct two consecutive minimization of $f(v,\bm{\phi})$  via  a quasi-Newton algorithm initiating at the coarse estimations.}

%------------------------------ Coarse  v_phi algorithm---------------
\begin{algorithm}[h]
	\caption{\textit{Coarse\_Velocity\_Angle\_Est}($\bm{z}_{\mathrm{\phi}},\{\bm{b}_k\}$)}\label{alg:coarse_vA_phi}
	\textbf{Inputs:}  Signal ($\bm{z}_{\mathrm{\phi}}\in \mathbb{C}^{L/2}$)\\
	\textbf{Parameters:}  DFT dimensions ($N_{\mathrm{\nu}}$), set of candidate \acp{aod} $\{\bm{\phi}_s\}_{s=0}^{N_{\mathrm{\phi}}-1}$ \\
	\textbf{Output:} $\hat{\bm{\phi}}$ and $\hat{v}_{\mathrm{r}}$
	\begin{algorithmic}[1]	
	\State $ \bm{F} \gets  N_{\mathrm{\nu}} \times L/2$ DFT matrix \label{CoarseVA_Line1}
	\For{$s\in\{0,\dots,N_{\mathrm{\phi}}-1\}$ }\label{CoarseVA_Line2}
	\For{$k\in\{0,\dots,L/2\}$ }\label{CoarseVA_Line3}
	\State $[\bm{z}_s]_{k} = \bm{a}(\theta)^{\top} \diag(\bm{b}_{k}) \bm{a}(\bm{\phi}_s)$\label{CoarseVA_Line4}
	\EndFor
	\State $\bm{w}_{s} = \bm{z}_{s}/\Vert \bm{z}_s \Vert$\label{CoarseVA_Line5}
	\State $\bm{g}_s = \bm{w}_{s}^{*} \odot \bm{z}_{\mathrm{\phi}}$\label{CoarseVA_Line6}
	\State $\bm{h}_s = \bm{F} \bm{g}_s$\label{CoarseVA_Line7}
	\EndFor
	\State $[i_{\mathrm{m}},s_{\mathrm{m}}] \gets \max_{i,s} \vert[\bm{h}_s]_i\vert$\label{CoarseVA_Line8}
	\State $\hat{\bm{\phi}} \gets \bm{\phi}_{s_{\mathrm{m}}}$\label{CoarseVA_Line9}
	\If{$i_{\mathrm{m}}> N_{\mathrm{\nu}}/2$}\label{CoarseVA_Line10}
	\State $i_{\mathrm{m}} \gets i_{\mathrm{m}} - N_{\mathrm{\nu}}+1$\label{CoarseVA_Line11}
	\EndIf
	\State $\hat{v}_{\mathrm{r}} \gets i_{\mathrm{m}}\lambda/(2T_{\mathrm{sym}}N_{\mathrm{\nu}})$\label{CoarseVA_Line12}\\
	\Return $\hat{\bm{\phi}}$ and $\hat{v}_{\mathrm{r}}$
	\end{algorithmic}
\end{algorithm}
%------------------------------end of Coarse  v_phi  algorithm---------------

\subsection{Estimation of channel parameters for the direct path}
Algorithm\,\ref{alg:estimator_direct} presents the estimation of channel parameters for the direct path based on some of the previous algorithms. The input is the received \ac{ofdm} signal $\bm{Y}$. First $\bm{w}_{\mathrm{b}}$ is used to extract the direct signal. Next, we estimate the value of $v_{\mathrm{b}}$, which requires solving a non-convex optimization. We solve this problem by first obtaining a coarse estimation in Line\,\ref{line3d:Est}.  \rev{In Line\,\ref{line4d:Est}, we use the \emph{refinement} step described in Sec.\,\ref{sec:estimation_v} using $\hat{v}_{\mathrm{b}}$ as the initial value}. Then, the effects of \ac{ue} mobility on the direct signal are compensated for and once again the direct signal is extracted via the vector $\bm{w}_{\mathrm{b}}$ in Line\,\ref{line6d:Est}. By compensating for the effects of \ac{ue} mobility, we reduce the residual interference and energy loss in  the \emph{matching} process (see Section\,\ref{sec:RisPhaseDesign}). One can see that the matrix $\hat{\bm{Z}}_{\mathrm{b}}$ in Line\,\ref{line6d:Est} is an estimate of ${\bm{Z}}_{\mathrm{b}}$ in \eqref{eq:Zbs}. The  matrix $\hat{\bm{Z}}_{\mathrm{b}}$ is then summed across time to establish $\bm{z}_{\tau}\in \mathbb{C}^N$. One can see that $\bm{z}_{\tau}$ has the structure $Lg_{\mathrm{b}} \left[\bm{D}(\tau_{\mathrm{b}})\right]_{:,1}$. Therefore, we use $\bm{z}_{\tau}$ to estimate $\tau_{\mathrm{b}}$ and then we use $\hat{\tau}_{\mathrm{b}}$ and $\bm{z}_{\tau}$ to estimate $g_{\mathrm{b}}$ in Line\,\ref{line10d:Est}.

%------------------------------main_Direct algorithm---------------
\begin{algorithm}[t]
	\caption{\textit{Direct\_Par\_Est}($\bm{Y}$) }\label{alg:estimator_direct}
	\textbf{Inputs:} Signal ($\bm{Y}\in \complexset{N}{L}$)\\
	\textbf{Output:} Estimation of parameters for the direct path: gain $\hat{g}_{\mathrm{b}}$,  radial velocity $\hat{v}_{\mathrm{b}}$, and delay $\hat{\tau}_{\mathrm{b}}$
	\begin{algorithmic}[1]
		\State $ \bm{w}_{\mathrm{b}} \gets [1,1]^\top$\label{line1d:Est}
		\State $ \bm{Z}_{\mathrm{b}} \gets \textit{match}(\bm{Y},\bm{w}_{\mathrm{b}})$\label{line2d:Est}
		\State $ \hat{v}_b \gets \textit{Coarse\_Velocity\_Est}(\bm{Z}_{\mathrm{b}})$\label{line3d:Est}
		\State $ \hat{v}_b \gets \textit{Fine\_Velocity\_Est}(\bm{Z}_{\mathrm{b}},\hat{v}_b)$\label{line4d:Est}
		%\State Calculate $\bm{C}(\hat{v}_b)$ and $\bm{E}(\hat{v}_b)$ based on \eqref{eq:Cmatrix} and \eqref{eq:Ematrix}, resp.
		\State  $\bm{T}_{\mathrm{b}} \gets \left(\bm{F}\bm{E}(\hat{v}_{\mathrm{b}})^{-1}\bm{F}^{\herm}\bm{Y}\right)\odot \bm{C}(\hat{v}_{\mathrm{b}})^*$\label{line5d:Est}
		\State  $\hat{\bm{Z}}_{\mathrm{b}} \gets \textit{match}(\bm{T}_{\mathrm{b}} ,\bm{w}_{\mathrm{b}})$\label{line6d:Est}
		\State $\bm{z}_{\tau} \gets \sum_t [\hat{\bm{Z}}_{\mathrm{b}}]_{:,t}$\label{line7d:Est}
		\State $\hat{\tau}_{\mathrm{b}}\gets\textit{Coarse\_delay\_Est}(\bm{z}_{\tau})$\label{line8:Est}
		\State $\hat{\tau}_{\mathrm{b}}\gets\textit{Fine\_delay\_Est}(\bm{z}_{\tau},\hat{\tau}_{\mathrm{b}})$\label{line9d:Est}
		%\State Calculate $\bm{C}(\hat{v}_b)$, $\bm{E}(\hat{v}_b)$, $\bm{D}(\hat{\tau}_{\mathrm{b}})$ based on \eqref{eq:Cmatrix}, \eqref{eq:Ematrix}, and \eqref{eq:matrixD} resp.
		\State $\hat{g}_{\mathrm{b}}\gets \left[\bm{D}(\hat{\tau}_{\mathrm{b}})\right]_{:,1}^{\herm}\bm{z}_{\tau}/(NL)$\label{line10d:Est}\\
		\Return $\hat{g}_{\mathrm{b}}, \hat{v}_{\mathrm{b}},$ and $ \hat{\tau}_{\mathrm{b}}$ \label{line11d:Est}
	\end{algorithmic}
\end{algorithm}
%------------------------------end of main direct algorithm---------------

\subsection{Estimation of channel parameters for the reflected path}\label{sec:Coars_Est_reflected_ch_par}
In this section, we use some of the previous algorithms to estimate the channel parameters for the reflected path. The process is described in Algorithm\,\ref{alg:estimator_reflected}. The input matrix $\hat{\bm{Y}}_{\mathrm{r}}$ is an estimate of ${\bm{Y}}_{\mathrm{r}}$ in \eqref{eq:Yr}. First, we match the signal with the vector $\bm{w}_{\mathrm{r}}$ to reduce the temporal dimension of the input signal from $L$ to $L/2$, without loss of information and also to remove any interference from the direct path (or possible scatterers). Then, we estimate  $\tau_{\mathrm{r}}$ in Line\,\ref{line3r:Est} and then compensate for its effects in Line\,\ref{line4r:Est}.
Next, we neglect the spatial-WB effects and assume that the effects of $v_{\mathrm{r}}$, and $\vectt{\phi}_{\mathrm{r}}$ are constant across the subcarriers. Therefore, to estimate these parameters we perform a summation across all subcarriers to obtain $\bm{z}_{\bm{\phi}}\in \mathbb{C}^{L/2}$. Assuming the spatial-NB model (see Section~\ref{sec:signalTransmissionNB}), one can see that $\bm{z}_{\bm{\phi}}$ has the structure at the \ac{rhs} of \eqref{eq:zPhi2}.
Therefore, we use $\bm{z}_{\bm{\phi}}$ to estimate $v_{\mathrm{r}}$ and $\vectt{\phi}_{\mathrm{r}}$ jointly in Line\,\ref{line6r:Est}. %We note that neglecting the special-WB effects in this step allows us to reduce the signal dimension by a factor of $N$, which simplifies the estimator and reduces its complexity.
 Next, we compensate for the effects of \ac{ue} mobility via $\hat{v}_{\mathrm{r}}$ in Line\,\ref{line8r:Est}. This reduces the interpath interference and the energy loss due to the \ac{ue} mobility. The steps from Line\,\ref{line9r:Est} to Line\,\ref{line13r:Est} refine the estimations obtained in Lines\,\ref{line2r:Est}--\ref{line6r:Est} \rev{(using the corresponding \emph{refinement} steps in Sec.\ref{sec:estimation_v}--\ref{sec:Estimation_va})\footnote{\rev{The extra inputs serve as initial values for the quasi-Newton algorithm. Specifically $\textit{Fine\_Velocity\_Angle\_Est}(\bm{z}_{\mathrm{\phi}}, 0, \hat{\bm{\phi}}, \{\bm{b}_k\})$ uses the initial values $0$ and $\hat{\bm{\phi}}$ for searching along the velocity and \ac{aod} dimensions, respectively.}}}. However, since the effects of velocity have been already compensated for in Line~\ref{line8r:Est},  we estimate the residual velocity ${\Delta}\hat{v}$ (with zero as the initial estimate) in Line\,\ref{line13r:Est}, which is then added  to the coarse estimation. This process is repeated $N_{\mathrm{iter}}$ times to obtain an accurate estimation, where $N_{\mathrm{iter}}$ is a design parameter. \rev{Alternatively, one can also stop the iterations after the difference between the estimated velocities becomes less than a certain threshold. }

%------------------------------main reflected algorithm---------------
\begin{algorithm}[t]
	\caption{\textit{Reflected\_Par\_Est}($\hat{\bm{Y}}_{\mathrm{r}}, \{\bm{b}_0,\dots \bm{b}_{L/2-1}$\}) }\label{alg:estimator_reflected}
	\textbf{Inputs:} Signal ($\hat{\bm{Y}}_{\mathrm{r}} \in \mathbb{C}^{N\times L}$), beams $\{\bm{b}_k\}$\\
	\textbf{Parameters:} Number of iterations $N_{\mathrm{iter}}$  \\
	\textbf{Output:} Estimation of parameters for the reflected path: AoD $\hat{\phi}$,  radial velocity $\hat{v}_{\mathrm{r}}$,  delay $\hat{\tau}_{\mathrm{r}}$
	\begin{algorithmic}[1]
		\State $ \bm{w}_{\mathrm{r}} \gets [1,-1]^\top$\label{line1r:Est}
		\State $ \hat{\bm{Z}}_{\mathrm{r}} \gets \textit{match}(\hat{\bm{Y}}_{\mathrm{r}},\bm{w}_{\mathrm{r}})$\label{line2r:Est}
		\State $\hat{\tau}_{\mathrm{r}}\gets\textit{Coarse\_delay\_Est}(\hat{\bm{Z}}_{\mathrm{r}})$\label{line3r:Est}
		\State $ \bm{T}_{\mathrm{r}} \gets \hat{\bm{Z}}_{\mathrm{r}}\odot [D(\hat{\tau}_{\mathrm{r}})^*]_{:,0:L/2-1}$\label{line4r:Est}
		\State $\bm{z}_{\mathrm{\phi}} \gets \sum_n [\bm{T}_{\mathrm{r}}]_{n,:}^{\top}$\label{line5r:Est}
		\State $[\hat{\phi}, \hat{v}_{\mathrm{r}}]\gets\textit{Coarse\_Velocity\_Angle\_Est}(\bm{z}_{\mathrm{\phi}},\{\bm{b}_k\})$\label{line6r:Est}
		\For{$i\in\{0,\dots, N_{\mathrm{iter}}\}$}\label{line7r:Est}
		\State  $\hat{\bm{Y}}_{\mathrm{rs}} \gets \left(\bm{F}\bm{E}(\hat{v}_{\mathrm{r}})^{-1}\bm{F}^{\herm}\hat{\bm{Y}}_{\mathrm{r}}\right)\odot \bm{C}(\hat{v}_{\mathrm{r}})^*$\label{line8r:Est}
		\State $ \hat{\bm{Z}}_{\mathrm{rs}}\gets \textit{match}(\hat{\bm{Y}}_{\mathrm{rs}},\bm{w}_{\mathrm{r}})$\label{line9r:Est}
		\State $\hat{\tau}_{\mathrm{r}}\gets\textit{Fine\_delay\_Est}(\hat{\bm{Z}}_{\mathrm{rs}},\hat{\tau}_{\mathrm{r}})$\label{line10r:Est}
		\State $ \bm{T}_{\mathrm{rs}} \gets \hat{\bm{Z}}_{\mathrm{rs}}\odot [D(\hat{\tau}_{\mathrm{r}})^*]_{:,0:L/2-1}$\label{line11r:Est}
		\State $\bm{z}_{\mathrm{\phi}} \gets \sum_n [\bm{T}_{\mathrm{rs}}]_{n,:}$\label{line12r:Est}
		\State $[\Delta\hat{v}_{\mathrm{r}}$, $\hat{\bm{\phi}}]\gets\textit{Fine\_Velocity\_Angle\_Est}(\bm{z}_{\mathrm{\phi}}, 0, \hat{\bm{\phi}}, \{\bm{b}_k\})$\label{line13r:Est}
		\State $\hat{v}_{\mathrm{r}} = \hat{v}_{\mathrm{r}}+\Delta\hat{v}_{\mathrm{r}}$\label{line14r:Est}
		\EndFor\\
		\Return $\hat{\phi}, \hat{v}_{\mathrm{r}},$ and $ \hat{\tau}_{\mathrm{r}}$ \label{line15r:Est}
	\end{algorithmic}
\end{algorithm}
%------------------------------end of main reflected algorithm---------------

\subsection{Estimation of UE position}
Algorithm\,\ref{alg:pos} explains how to estimate the position of the \ac{ue}  via geometrical channel parameters. First, we calculate the direction of the \ac{ue} based on $\hat{\bm{\phi}}$ in Line\,\ref{line2:pos}. Next, based on \eqref{eq:taub} and \eqref{eq:taur}, we can estimate the distance between the \ac{ue} and the RIS by minimizing the function $f(d)$, defined in Line\,\ref{line3:pos}. 
%------------------------------ Pos Est algorithm---------------
\begin{algorithm}[h]
	\caption{$\textit{Position\_Est}(\hat{\tau}_{\mathrm{b}},\hat{\tau}_{\mathrm{r}},\hat{\bm{\phi}})$ }\label{alg:pos}
	\textbf{Inputs:} Estimation of the \acp{toa} ($\hat{\tau}_{\mathrm{b}}$,$\hat{\tau}_{\mathrm{r}}$) and \ac{aod}  ($\hat{\bm{\phi}}$) \\
	\textbf{Output:} $\hat{\bm{p}}$
	\begin{algorithmic}[1]
		\State $\Delta r\gets c\vert\hat{\tau}_{\mathrm{r}}-\hat{\tau}_{\mathrm{b}}\vert$\label{line1:pos}
		\State $\bm{k}\gets \begin{bmatrix}
\sin([\bm{\hat{\phi}}]_{\mathrm{el}})\cos([\bm{\hat{\phi}}]_{\mathrm{az}})\\
\sin([\bm{\hat{\phi}}]_{\mathrm{el}})\sin([\bm{\hat{\phi}}]_{\mathrm{az}})\\
\cos([\bm{\hat{\phi}}]_{\mathrm{el}})
\end{bmatrix}$\label{line2:pos}
		\State $f(d)\gets \left(d+\Vert\bm{p}_{\mathrm{b}}-\bm{p}_{\mathrm{r}}\Vert-\Vert\bm{p}_{\mathrm{b}}-\bm{p}_{\mathrm{r}}-d\bm{k}\Vert-\Delta r\right)^2$\label{line3:pos}
		\State $d_{\mathrm{m}}\gets \min_df(d)$ \label{line4:pos}
		\State $\hat{\bm{p}}\gets d_{\mathrm{m}}\bm{k}$\label{line5:pos}\\
		\Return $\hat{\bm{p}}$\label{line6:pos}
	\end{algorithmic}
\end{algorithm}

\subsection{Overall process}
The overall estimation process is described in Algorithm~\ref{alg:estimator}. First, the direct channel parameters are estimated in Line\,\ref{line1:Est}.
Next, we obtain an estimation of the direct signal and remove it from the received signal to obtain an estimate of the reflected one, which is used to obtain estimates of the reflected channel parameters.  Next, we use the estimate of the geometrical channel parameters to find the \ac{ue} position. Finally, in Line\,\ref{line5:Est}, we use \eqref{eq:taub} to estimate the \ac{ue} clock bias.

\vspace{.5cm}
\paragraph*{Complexity}
Algorithm\,\ref{alg:estimator} has a low complexity compared to a search over all the possible values of the channel parameters to maximize the likelihood function, which requires a 6-dimensional search (the optimal values of the gains can be calculated in closed-form). Note that our estimator performs at most a \rev{3}-dimensional search at each step. \rev{Specifically,  Algorithms\,\ref{alg:coarse_v} and \ref{alg:coarse_tau} (and their corresponding refinement step) each apply only one line search each to find an estimation of the radial velocity and delay, respectively.  Algorithm\,\ref{alg:coarse_vA_phi} searches over the possible radial velocities and also the AoDs (elevations and azimuths), consequently, the search is performed over a 3D space, while its corresponding refinement step applies a 1D and a 2D search. Furthermore, we significantly reduced the complexity of our 3D search by using FFT for searching over velocities and  2D FFT method to search over possible AoDs (see Appendix\,\ref{app:fft}). 
Therefore, the proposed algorithm} requires  much less computational power compared to the maximum-likelihood estimator.

%------------------------------main algorithm---------------
\begin{algorithm}[t]
	\caption{\textit{Estimator}($\bm{Y}, \{\bm{b}_0,\dots \bm{b}_{L/2-1}$\}) }\label{alg:estimator}
	\textbf{Inputs:} Received signal ($\bm{Y}\in \complexset{N}{L}$), beams $\{\bm{b}_k\}$\\
	\textbf{Output:} Estimation of UE position ($\hat{\bm{p}}$),  UE clock bias $\hat{\Delta}_t$, and radial velocities $\hat{v}_{\mathrm{b}}, \hat{v}_{\mathrm{r}}$ 
	\begin{algorithmic}[1]
		\State $[\hat{g}_{\mathrm{b}}, \hat{v}_{\mathrm{b}}, \hat{\tau}_{\mathrm{b}}]\gets$\textit{Direct\_Par\_Est}($\bm{Y}$)\label{line1:Est}
		\State $\hat{\bm{Y}}_{\mathrm{r}}\gets \bm{Y} - \hat{g}_{\mathrm{b}} \bm{F}\bm{E}(\hat{v}_{\mathrm{b}}) \bm{F}^{\herm} \left(\bm{D}(\hat{\tau}_{\mathrm{b}}) \odot \bm{C}(\hat{v}_{\mathrm{b}})\right)$\label{line2:Est}
		\State $[\hat{\phi}, \hat{v}_{\mathrm{r}}, \hat{\tau}_{\mathrm{r}}]\gets$\textit{Reflected\_Par\_Est}($\hat{\bm{Y}}_{\mathrm{r}}, \{\bm{b}_0,\dots \bm{b}_{L/2-1}$\})\label{line3:Est}
		\State $\hat{\bm{p}}\gets \textit{Position\_Est}(\hat{\tau}_{\mathrm{b}},\hat{\tau}_{\mathrm{r}},\hat{\bm{\phi}})$\label{line4:Est}
		\State $\hat{\Delta}_t \gets \hat{\tau}_{\mathrm{b}}-\Vert\hat{\bm{p}}-\bm{p}_{\mathrm{b}}\Vert/c$\label{line5:Est}\\
		\Return $\hat{\bm{p}}$, $\hat{\Delta}_t$, $\hat{v}_{\mathrm{b}}$, and $\hat{v}_{\mathrm{r}}$ \label{line6:Est}
	\end{algorithmic}
\end{algorithm}
%------------------------------end of main algorithm---------------

%------------------------------Pos Est algorithm---------------
\section{Simulation results}\label{sec:simulationResults}

%-----------table-----------------

\begin{table}[!t]
\vspace{.1cm}
	\caption{Parameters used in the simulation.}
	\label{table:par}
	\centering
	\begin{tabular}{l l l }
		\hline
		\hline
		Parameter&Symbol& Value\\
		\hline
		RIS dimensions & $M_1\times M_2$ &$64\times 64$\\
		Wavelength & $\lambda$ & $1 \ {\mathrm{cm}}$\\
		RIS element distance & $d$ &$0.5 \ {\mathrm{cm}}$\\
		Light speed & $c$ & $3\times 10^8 \ \mathrm{m/s}$\\
		Number of subcarriers & $\numSubCarriers$ & $3\, 000$\\
		Subcarrier bandwidth & $\deltaf$ & $120 \ \mr{kHz}$\\
		Symbol duration & $T$ & $8.33 \ \mr{us}$\\
		CP duration & $\Tcp$ & $0.58 \ \mr{us}$\\
		Number of transmissions & $L$ & $256$\\
		Transmission Power &$N \deltaf E_{\mathrm{s}}$ & $20 \ \mathrm{dBm}$\\
		Noise PSD & $N_0$ & $-174 \ \mathrm{dBm/Hz}$\\
		UE's Noise figure& $n_f$ & $8 \ \mathrm{dB}$\\
		Noise variance& $\sigma^2=n_f N_0$ & $-166 \ \mathrm{dBm/Hz}$\\
		BS position & $\bm{p}_{\mathrm{b}}$ & $[5,5,0]$\\
		RIS position & $\bm{p}_{\mathrm{r}}$ & $[0,0,0]$\\
		Uncertainty radius & $ \sigma$ & $1$m\\
		\hline
		\hline
	\end{tabular}
\end{table}

%-------------
In this section, we assess the accuracy of our estimation method and compare it to the \ac{crb} for a system example with default parameters listed in Table\,\ref{table:par}. The algorithm parameters are set to $N_{\mathrm{\tau}} = 4096$ \rev{(the IDFT dimension for delay estimation in Algorithm\,\ref{alg:coarse_tau})} and $N_{\mathrm{v}} = N_{\mathrm{\nu}} = 256 $ \rev{(the DFT dimension for velocity estimation in Algorithms\,\ref{alg:coarse_tau} and \ref{alg:coarse_vA_phi}, respectively)}. \rev{The number of candidate \acp{aod} in Algorithm\,\ref{alg:coarse_vA_phi} is set to}   $N_{\mathrm{\phi}} = 256$ when using the directional profiles or $N_{\mathrm{\phi}} = 256^2$ for the random profiles, \rev{and the selection of candidate \acp{aod} are done according to Appendix\,\ref{app:fft}.} \rev{Also, the number of iterations in Algorithm\,\ref{alg:estimator_reflected} is set to}\footnote{\rev{Based on our simulation results (not provided in this paper), for the considered parameters, the position error saturates after two iterations.}} $N_{\mathrm{itr}} = 3$. The \ac{ris} is located at the origin such that the local coordinates of RIS matches the global coordinate system ($\bm{R}$ is the identity matrix). \rev{Following the widely used assumption of quasi-static channel over a coherence interval\rev{\footnote{\rev{The UE mobility affects the time-varying phase of the received signal through Doppler-induced phase progressions in fast-time and slow-time domains, modeled by \eqref{eq:Ematrix} and \eqref{eq:Cmatrix}, respectively.}}} \cite{IRS_OFDM_TCOM_2020,Zhang_RIS_CE_2019,adaptiveOFDM_RIS_JSAC_2020,CE_OFDM_WCL_2021,CE_practical_2021,OFDMA_IRS_WCL_2020} (consisting of $L$ OFDM symbols)}, the channel gains are assigned random phases \rev{(fixed during $L$ symbols)} and the amplitudes are calculated as \cite[Eq. (21)--(23)]{ellingson2019path}
\begin{align}
    \vert g_{\mathrm{b}}\vert &= \frac{\lambda\sqrt{E_{\mathrm{s}}}}{4\pi\Vert\bm{p}_{\mathrm{b}}-\bm{p}\Vert}\\
 \vert g_{\mathrm{r}}\vert &= \frac{\lambda^2 \cos^q(\alpha_{\mathrm{\theta}})\cos^q(\alpha_{\mathrm{\phi}})\sqrt{E_{\mathrm{s}}}}{16\pi\Vert\bm{p}_{\mathrm{b}}-\bm{p}_{\mathrm{r}}\Vert\Vert\bm{p}_{\mathrm{r}}-\bm{p}\Vert}
\end{align}
with $q=0.285$ \rev{(see \cite{ellingson2019path})}, where $E_{\mathrm{s}}$ indicates the pilot energy, and $\alpha_{\mathrm{\phi}}$ and $\alpha_{\mathrm{\theta}}$ are defined below \eqref{eq:condSpWB}.
Before presenting the results, in Section\,\ref{sec:fimAnalysis}, we present some preliminary information about the calculation of the \ac{crb} based on \ac{fim} analysis, which will be used as a benchmark in this section.
Then, we study the spatial-WB effects in Section\,\ref{sec:Results_wbEffects} for different RIS sizes and signal bandwidths. Next in Section\,\ref{sec:Results_mobilityEffects} the mobility effects are considered and the influence of the uncertainty radius as well as scatterers are shown in Section\,\ref{sec:radius}.

\subsection{FIM analysis}\label{sec:fimAnalysis}
 \ac{fim} analysis can be used to develop  theoretical lower bounds on the estimation error of any unbiased estimator. We do so by calculating the \ac{fim} first for the channel parameters and then for the positional parameters. We define the set of channel parameters as
\begin{align}
    \bm{\zeta}_{\mathrm{ch}} = [&  \tau_{\mathrm{b}} , \tau_{\mathrm{r}}, [\bm{\phi}]_{\mathrm{az}}, [\bm{\phi}]_{\mathrm{el}}, v_{\mathrm{b}}, v_{\mathrm{r}}, 
    \Re(g_{\mathrm{b}}), \Im(g_{\mathrm{b}}), \Re(g_{\mathrm{r}}), \Im(g_{\mathrm{r}}) ]^\top.
\end{align}
The \ac{fim} can be calculated as follows \cite{kayEstimation}
\begin{align}
    \fisherInfoch = \frac{2}{\sigma^2}\sum\limits_{t=0}^{L-1}\sum\limits_{n=0}^{\numSubCarriers-1}\realPart\left\{\frac{\rond [\meanSig]_{n,t}}{\rond \parameterCh} \left(\frac{\rond [\meanSig]_{n,t}}{\rond \parameterCh}\right)^\herm\right\},\label{eq:fimch}
\end{align}
where $\meanSig$ is the noiseless part of the received signal. \rev{In this paper, we use the dynamic spatial-wideband model in  \eqref{eq:Yb}--\eqref{eq:Yr} to compute \eqref{eq:fimch} unless stated otherwise.}
%We calculate the derivatives associated with \eqref{eq:fimch} in Appendix~\ref{app:chFIM}. 
Next, we calculate the \ac{fim} for positional parameters, that is
\begin{align}
    \bm{\zeta}_{\mathrm{po}} = [\bm{p}^\top, \Delta t, v_{\mathrm{b}}, v_{\mathrm{r}}, 
    \Re(g_{\mathrm{b}}), \Im(g_{\mathrm{b}}), \Re(g_{\mathrm{r}}), \Im(g_{\mathrm{r}})]^\top.
\end{align}
We do so, by calculating $\fisherInfpo = \jacob^\top \fisherInfoch \jacob$, where the Jacobian matrix $\jacob\in\realSet^{10\times 10}$ is defined as
\begin{align}
    \jacob_{\ell,s} = \frac{\rond [\parameterCh]_\ell}{\rond [\parameterPo]_s}.\label{eq:jacobElements}
\end{align}
%we calculate the elements of $\jacob$ in Appendix~\ref{app:jacob}.
By obtaining $\fisherInfch$ the estimation error of the $m$th channel parameter is lower bounded as
\begin{align}
    \mathrm{E}(\vert [\bm{\zeta}_{\mathrm{ch}}]_m-\widehat{[\bm{\zeta}_{\mathrm{ch}}]}_m\vert^2)\geq [\fisherInfch^{-1}]_{m,m},
\end{align}
where $\widehat{[\bm{\zeta}_{\mathrm{ch}}]}_m$ indicates the estimate of the parameter $[\bm{\zeta}_{\mathrm{ch}}]_m$.
Similarly the estimation of the positional parameters can be bounded using $\fisherInfpo$. Furthermore, we use the \ac{peb}  as a lower bound on the position estimation error, that is 
\begin{align}
    \sqrt{\big[\mathrm{E}(\Vert \bm{p}-\hat{\bm{p}}\Vert^2)\big]}\geq \sqrt{\mathrm{trace}([\fisherInfpo^{-1}]_{1:3,1:3})}.
\end{align}
The derivatives required for calculating  $\fisherInfch$ and $\jacob$ can be calculated based on the relations described in Section\,\ref{sec:systemModChannelModel}. 

\rev{We note that for the directional codebook, the prior information of the UE position affects the FIM only through the beamforming, and we do not take into account the effects of the fusion of the estimated position and the prior information. Since the proposed estimator also does not perform information fusion, the presented PEB correctly lower-bounds the position error of our estimator.}

\begin{figure}
    \centering
    \includegraphics[width =5cm]{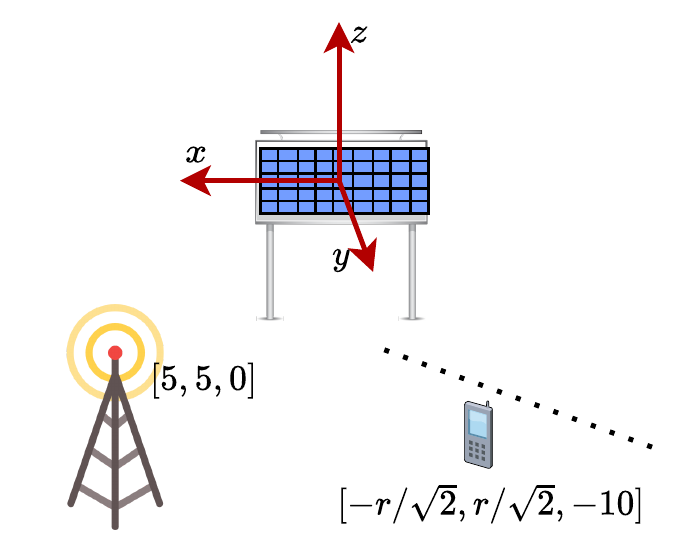}
    \caption{\rev{Placement of the RIS, BS, and UE in the 3D space.}}
    \label{fig:struct}
\end{figure}

%-------------  figure---------------
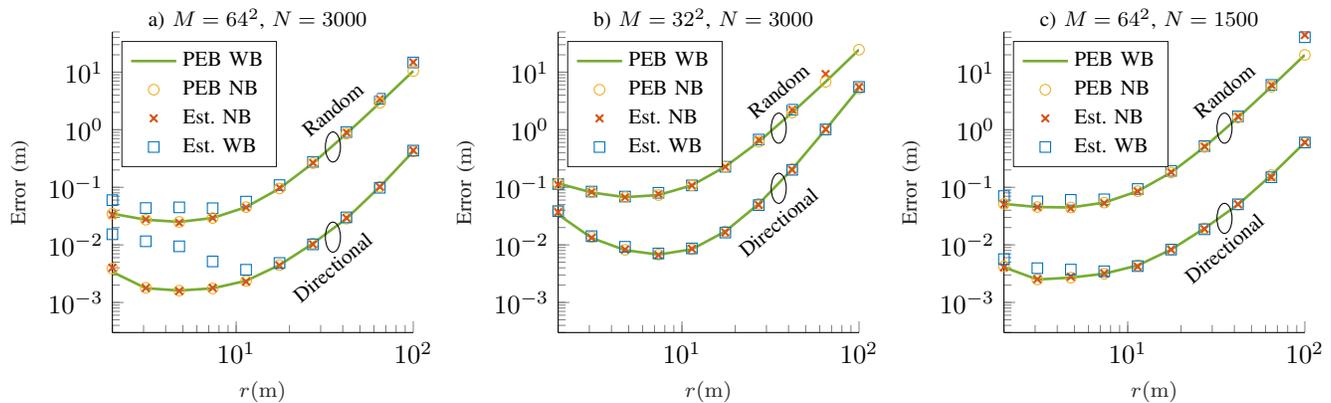
\begin{figure*}[t]
    \centering
    \begin{subfigure}[b]{0.32\textwidth}
\definecolor{mycolor1}{rgb}{0.00000,0.44700,0.74100}%
\definecolor{mycolor2}{rgb}{0.85000,0.32500,0.09800}%
\definecolor{mycolor3}{rgb}{0.92900,0.69400,0.12500}%
\definecolor{mycolor4}{rgb}{0.49400,0.18400,0.55600}%
\definecolor{mycolor5}{rgb}{0.46600,0.67400,0.18800}%
\definecolor{mycolor6}{rgb}{0.30100,0.74500,0.93300}%
\definecolor{mycolor7}{rgb}{0.63500,0.07800,0.18400}%
\begin{tikzpicture}

\begin{axis}[%
width=4cm,
height=4cm,
at={(1.011in,0.642in)},
scale only axis,
xmin=2,
xmax=100,
xlabel style={font=\color{white!15!black}},
xlabel={\footnotesize $r(\mr{m})$},
ymode=log,
xmode=log,
ymin=3e-4,
ymax=50,
yminorticks=true,
ylabel style={font=\color{white!15!black}},
ylabel={\footnotesize  Error ($\mr{m}$)},
axis background/.style={fill=white},
title style={font=\bfseries},
axis x line*=bottom,
axis y line*=left,
legend style={legend cell align=left, align=left, draw=white!15!black},
legend pos=north west
]
\addplot [color=mycolor5, line width=1.0pt]
  table[row sep=crcr]{%
2	0.00338761899889947\\
3.08890420989276	0.00176938070385056\\
4.7706646089466	0.00161025343767585\\
7.36806299728077	0.001758585223333\\
11.3796204055278	0.00237612136800012\\
17.5752786888082	0.00446797855939566\\
27.1441761659491	0.0105772225131485\\
41.9228800165354	0.0285536572088744\\
64.7478802869525	0.101854670946259\\
100	0.420739696988352\\
};
\addlegendentry{\footnotesize PEB WB}

\addplot [color=mycolor6, only marks, mark size=2.0pt, mark=o, mark options={solid, mycolor3}]
  table[row sep=crcr]{%
2	0.00386660767904022\\
3.08890420989276	0.00179699360126705\\
4.7706646089466	0.00160445891419049\\
7.36806299728077	0.00175014369767319\\
11.3796204055278	0.00236983410134175\\
17.5752786888082	0.00446061670399269\\
27.1441761659491	0.0105640385852201\\
41.9228800165354	0.0285365237597888\\
64.7478802869525	0.101823728962725\\
100	0.420679478683035\\
};
\addlegendentry{\footnotesize PEB NB}

\addplot [color=mycolor2, only marks, line width=0.8pt,mark size=2.0pt, mark=x, mark options={solid, mycolor2}]
  table[row sep=crcr]{%
2	0.00401355997489908\\
3.08890420989276	0.00178484986887984\\
4.7706646089466	0.00158081678205376\\
7.36806299728077	0.00179191873090889\\
11.3796204055278	0.00229595106642901\\
17.5752786888082	0.00439365233187191\\
27.1441761659491	0.0101817546941529\\
41.9228800165354	0.0293876229880932\\
64.7478802869525	0.102178181201358\\
100	0.437507813575822\\
};
\addlegendentry{\footnotesize Est. NB}

\addplot [color=mycolor1, only marks, mark size=2.0pt, mark=square, mark options={solid, mycolor1}]
  table[row sep=crcr]{%
2	0.0152721013919874\\
3.08890420989276	0.0115449943507107\\
4.7706646089466	0.00944255526212406\\
7.36806299728077	0.0051581422921131\\
11.3796204055278	0.00369778577573066\\
17.5752786888082	0.00485140839431361\\
27.1441761659491	0.0101971234187339\\
41.9228800165354	0.0300566613149074\\
64.7478802869525	0.0981218719257098\\
100	0.434273810152172\\
};
\addlegendentry{\footnotesize Est. WB}

\addplot [color=mycolor5, line width=1.0pt]
  table[row sep=crcr]{%
2	0.0348040147115256\\
3.08890420989276	0.0273332974642485\\
4.7706646089466	0.0250620405542816\\
7.36806299728077	0.0292705657923253\\
11.3796204055278	0.0452893859675634\\
17.5752786888082	0.095466574973231\\
27.1441761659491	0.262997900793525\\
41.9228800165354	0.851226441140257\\
64.7478802869525	2.92906324817106\\
100	10.4548623484736\\
};
%\addlegendentry{PEB RanWB}

\addplot [color=mycolor1, only marks, mark size=2.0pt, mark=square, mark options={solid, mycolor1}]
  table[row sep=crcr]{%
2	0.059816334925875\\
3.08890420989276	0.0436022869933928\\
4.7706646089466	0.0448324318505374\\
7.36806299728077	0.0432995440211496\\
11.3796204055278	0.0564308376594689\\
17.5752786888082	0.1091335004084\\
27.1441761659491	0.276192531219088\\
41.9228800165354	0.910038617363352\\
64.7478802869525	3.45017413800754\\
100	14.7168987832555\\
};
%\addlegendentry{Est RanWB}

\addplot [color=mycolor2, only marks,line width=0.8pt, mark size=2.0pt, mark=x, mark options={solid, mycolor2}]
  table[row sep=crcr]{%
2	0.033142222504042\\
3.08890420989276	0.0277866394044449\\
4.7706646089466	0.0241601747002731\\
7.36806299728077	0.0293953009390564\\
11.3796204055278	0.0446148004683759\\
17.5752786888082	0.0986839171628428\\
27.1441761659491	0.267269441996122\\
41.9228800165354	0.899379538534629\\
64.7478802869525	3.40787443306833\\
100	14.8540455643638\\
};
%\addlegendentry{Est RanNB}

\addplot [color=mycolor3, only marks, mark size=2.0pt, mark=o, mark options={solid, mycolor3}]
  table[row sep=crcr]{%
2	0.0348040147115256\\
3.08890420989276	0.0273332974642485\\
4.7706646089466	0.0250620405542816\\
7.36806299728077	0.0292705657923253\\
11.3796204055278	0.0452893859675634\\
17.5752786888082	0.095466574973231\\
27.1441761659491	0.262997900793525\\
41.9228800165354	0.851226441140257\\
64.7478802869525	2.92906324817106\\
100	10.4548623484736\\
};
%\addlegendentry{PEB RanWB}
\end{axis}
\node[rotate=40,fill=white] (BOC6) at (5.5cm,2.5cm){\footnotesize Directional};
\node[rotate=40,fill=white] (BOC6) at (5.5cm,4.6cm){\footnotesize Random};
\node[rotate=0,fill=white] (BOC6) at (4.5cm,5.8cm){\footnotesize a) $M=64^2$, $N=3000$};
\draw (5.5cm,4.1cm) ellipse (.1cm and .2cm);
\draw (5.5cm,2.9cm) ellipse (.1cm and .2cm);
\end{tikzpicture}%
    \end{subfigure}
    %\hspace{1cm}
    \begin{subfigure}[b]{0.32\textwidth}
\definecolor{mycolor1}{rgb}{0.00000,0.44700,0.74100}%
\definecolor{mycolor2}{rgb}{0.85000,0.32500,0.09800}%
\definecolor{mycolor3}{rgb}{0.92900,0.69400,0.12500}%
\definecolor{mycolor4}{rgb}{0.49400,0.18400,0.55600}%
\definecolor{mycolor5}{rgb}{0.46600,0.67400,0.18800}%
\definecolor{mycolor6}{rgb}{0.30100,0.74500,0.93300}%
\definecolor{mycolor7}{rgb}{0.63500,0.07800,0.18400}%
\begin{tikzpicture}

\begin{axis}[%
width=4cm,
height=4cm,
at={(1.011in,0.642in)},
scale only axis,
xmin=2,
xmax=100,
xlabel style={font=\color{white!15!black}},
xlabel={\footnotesize $r(\mr{m})$},
ymode=log,
xmode=log,
ymin=3e-4,
ymax=50,
yminorticks=true,
ylabel style={font=\color{white!15!black}},
ylabel={\footnotesize  Error ($\mr{m}$)},
axis background/.style={fill=white},
title style={font=\bfseries},
axis x line*=bottom,
axis y line*=left,
legend style={legend cell align=left, align=left, draw=white!15!black},
legend pos=north west
]
\addplot [color=mycolor5, line width=1.0pt]
  table[row sep=crcr]{%
2	0.0336480606456222\\
3.08890420989276	0.0132169384692285\\
4.7706646089466	0.00810559351272447\\
7.36806299728077	0.00677313501897633\\
11.3796204055278	0.0083868977553078\\
17.5752786888082	0.0171323541361305\\
27.1441761659491	0.0514956306382615\\
41.9228800165354	0.206323682519051\\
64.7478802869525	1.0090374900368\\
100	5.2105879805614\\
};
\addlegendentry{\footnotesize PEB WB}

\addplot [color=mycolor6, only marks, mark size=2.0pt, mark=o, mark options={solid, mycolor3}]
  table[row sep=crcr]{%
2	0.0389011289247915\\
3.08890420989276	0.0137363362234297\\
4.7706646089466	0.00821595495118925\\
7.36806299728077	0.00678540652847148\\
11.3796204055278	0.00838619627596608\\
17.5752786888082	0.01712926295951\\
27.1441761659491	0.051492942953474\\
41.9228800165354	0.206322085831634\\
64.7478802869525	1.00903523249723\\
100	5.21059711420372\\
};
\addlegendentry{\footnotesize PEB NB}

\addplot [color=mycolor2, only marks, line width=0.8pt,mark size=2.0pt, mark=x, mark options={solid, mycolor2}]
  table[row sep=crcr]{%
2	0.0369680129247881\\
3.08890420989276	0.0132626648909565\\
4.7706646089466	0.00830539951968002\\
7.36806299728077	0.00663590554664578\\
11.3796204055278	0.00852709344004224\\
17.5752786888082	0.0164165005238378\\
27.1441761659491	0.0493011168297599\\
41.9228800165354	0.201515021921401\\
64.7478802869525	1.02132106901445\\
100	5.53061593987541\\
};
\addlegendentry{\footnotesize Est. NB}

\addplot [color=mycolor1, only marks, mark size=2.0pt, mark=square, mark options={solid, mycolor1}]
  table[row sep=crcr]{%
2	0.0385556538565782\\
3.08890420989276	0.0141584387963908\\
4.7706646089466	0.00925941202908926\\
7.36806299728077	0.00709819323544725\\
11.3796204055278	0.00868623646285409\\
17.5752786888082	0.0163681007457189\\
27.1441761659491	0.0490729733977735\\
41.9228800165354	0.201094015329017\\
64.7478802869525	1.01214096921815\\
100	5.53961901805637\\
};
\addlegendentry{\footnotesize Est. WB}

\addplot [color=mycolor5, line width=1.0pt]
  table[row sep=crcr]{%
2	0.117108157979531\\
3.08890420989276	0.0812644222069627\\
4.7706646089466	0.0674587812143623\\
7.36806299728077	0.0733132969009929\\
11.3796204055278	0.108261208840832\\
17.5752786888082	0.228679999799574\\
27.1441761659491	0.618520942954977\\
41.9228800165354	1.9845431492406\\
64.7478802869525	6.79888918973137\\
100	24.5718864241235\\
};
%\addlegendentry{PEB RanWB}

\addplot [color=mycolor1, only marks, mark size=2.0pt, mark=square, mark options={solid, mycolor1}]
  table[row sep=crcr]{%
2	0.113747980688834\\
3.08890420989276	0.0844176872603005\\
4.7706646089466	0.0690702186853737\\
7.36806299728077	0.0802044849524576\\
11.3796204055278	0.107882111285784\\
17.5752786888082	0.227470687656198\\
27.1441761659491	0.674811340174259\\
41.9228800165354	2.22675623917591\\
64.7478802869525	4052.52430420624\\
100	221534.904467303\\
};
%\addlegendentry{Est RanWB}

\addplot [color=mycolor2, only marks,line width=0.8pt, mark size=2.0pt, mark=x, mark options={solid, mycolor2}]
  table[row sep=crcr]{%
2	0.110683279724819\\
3.08890420989276	0.0829946613677911\\
4.7706646089466	0.0669560134414416\\
7.36806299728077	0.0767071184469972\\
11.3796204055278	0.107286763920761\\
17.5752786888082	0.226560942739164\\
27.1441761659491	0.670801166210572\\
41.9228800165354	2.22215588612127\\
64.7478802869525	9.28729757785314\\
100	157787.925926084\\
};
%\addlegendentry{Est RanNB}

\addplot [color=mycolor3, only marks, mark size=2.0pt, mark=o, mark options={solid, mycolor3}]
  table[row sep=crcr]{%
2	0.117069254365652\\
3.08890420989276	0.0812546366464898\\
4.7706646089466	0.0673766187107247\\
7.36806299728077	0.0733025094986203\\
11.3796204055278	0.108271165910849\\
17.5752786888082	0.228730591688395\\
27.1441761659491	0.618797665548627\\
41.9228800165354	1.98430542919093\\
64.7478802869525	6.799158392377\\
100	24.5689166118973\\
};
%\addlegendentry{PEB RanNB}
\end{axis}
\node[rotate=40,fill=white] (BOC6) at (5.5cm,3.1cm){\footnotesize Directional};
\node[rotate=40,fill=white] (BOC6) at (5.5cm,4.85cm){\footnotesize Random};
\draw (5.5cm,4.35cm) ellipse (.1cm and .2cm);
\draw (5.5cm,3.55cm) ellipse (.1cm and .2cm);
\node[rotate=0,fill=white] (BOC6) at (4.5cm,5.8cm){\footnotesize b) $M=32^2$, $N=3000$};
\end{tikzpicture}%
    \end{subfigure}
    \begin{subfigure}[b]{0.32\textwidth}
\definecolor{mycolor1}{rgb}{0.00000,0.44700,0.74100}%
\definecolor{mycolor2}{rgb}{0.85000,0.32500,0.09800}%
\definecolor{mycolor3}{rgb}{0.92900,0.69400,0.12500}%
\definecolor{mycolor4}{rgb}{0.49400,0.18400,0.55600}%
\definecolor{mycolor5}{rgb}{0.46600,0.67400,0.18800}%
\definecolor{mycolor6}{rgb}{0.30100,0.74500,0.93300}%
\definecolor{mycolor7}{rgb}{0.63500,0.07800,0.18400}%
\begin{tikzpicture}

\begin{axis}[%
width=4cm,
height=4cm,
at={(1.011in,0.642in)},
scale only axis,
xmin=2,
xmax=100,
xlabel style={font=\color{white!15!black}},
xlabel={\footnotesize $r(\mr{m})$},
ymode=log,
xmode=log,
ymin=3e-4,
ymax=50,
yminorticks=true,
ylabel style={font=\color{white!15!black}},
ylabel={\footnotesize  Error ($\mr{m}$)},
axis background/.style={fill=white},
title style={font=\bfseries},
axis x line*=bottom,
axis y line*=left,
legend style={legend cell align=left, align=left, draw=white!15!black},
legend pos=north west
]
\addplot [color=mycolor5, line width=1.0pt]
  table[row sep=crcr]{%
2	0.004095261722583\\
3.08890420989276	0.00250689379655056\\
4.7706646089466	0.00269768539270624\\
7.36806299728077	0.00315377239628888\\
11.3796204055278	0.00439286077920255\\
17.5752786888082	0.00828615640248694\\
27.1441761659491	0.0193720327977398\\
41.9228800165354	0.0499198830435804\\
64.7478802869525	0.162835881852595\\
100	0.599441558701474\\
};
\addlegendentry{\footnotesize PEB WB}

\addplot [color=mycolor6, only marks, mark size=2.0pt, mark=o, mark options={solid, mycolor3}]
  table[row sep=crcr]{%
2	0.00422910017892016\\
3.08890420989276	0.00250802073395915\\
4.7706646089466	0.00269127982194255\\
7.36806299728077	0.00314893938305376\\
11.3796204055278	0.00438938878513408\\
17.5752786888082	0.00828217975837362\\
27.1441761659491	0.0193646880646537\\
41.9228800165354	0.0499105973250608\\
64.7478802869525	0.162820577002329\\
100	0.599417228218133\\
};
\addlegendentry{\footnotesize PEB NB}

\addplot [color=mycolor2, only marks, line width=0.8pt,mark size=2.0pt, mark=x, mark options={solid, mycolor2}]
  table[row sep=crcr]{%
2	0.00407175398922714\\
3.08890420989276	0.00253593355056102\\
4.7706646089466	0.00279453975465754\\
7.36806299728077	0.00319231863248556\\
11.3796204055278	0.00419470331408893\\
17.5752786888082	0.00828329983688311\\
27.1441761659491	0.0187671151850209\\
41.9228800165354	0.0508549688655933\\
64.7478802869525	0.149356833990212\\
100	0.603301864699532\\
};
\addlegendentry{\footnotesize Est. NB}

\addplot [color=mycolor1, only marks, mark size=2.0pt, mark=square, mark options={solid, mycolor1}]
  table[row sep=crcr]{%
2	0.00565633370964143\\
3.08890420989276	0.00393835708411887\\
4.7706646089466	0.00372200893701614\\
7.36806299728077	0.00347771490903331\\
11.3796204055278	0.00427620613842968\\
17.5752786888082	0.0082903441204218\\
27.1441761659491	0.0186415377002371\\
41.9228800165354	0.0506282665684905\\
64.7478802869525	0.149393535129891\\
100	0.600247032278519\\
};
\addlegendentry{\footnotesize Est. WB}

\addplot [color=mycolor5, line width=1.0pt]
  table[row sep=crcr]{%
2	0.0514937910720725\\
3.08890420989276	0.0454174593126512\\
4.7706646089466	0.0446477774624815\\
7.36806299728077	0.0542017694914833\\
11.3796204055278	0.0854914328484594\\
17.5752786888082	0.180647672296606\\
27.1441761659491	0.500798009387463\\
41.9228800165354	1.6223360725179\\
64.7478802869525	5.5769882605416\\
100	19.9297644252037\\
};
%\addlegendentry{PEB RanWB}

\addplot [color=mycolor1, only marks, mark size=2.0pt, mark=square, mark options={solid, mycolor1}]
  table[row sep=crcr]{%
2	0.0704487363379193\\
3.08890420989276	0.0569316992530344\\
4.7706646089466	0.0605412179307454\\
7.36806299728077	0.0618093967782262\\
11.3796204055278	0.0933791047542119\\
17.5752786888082	0.193116135348952\\
27.1441761659491	0.519755009344439\\
41.9228800165354	1.7047341453058\\
64.7478802869525	6.00720983383804\\
100	40.4246259441531\\
};
%\addlegendentry{Est RanWB}

\addplot [color=mycolor2, only marks,line width=0.8pt, mark size=2.0pt, mark=x, mark options={solid, mycolor2}]
  table[row sep=crcr]{%
2	0.0504050629053013\\
3.08890420989276	0.0453062396573417\\
4.7706646089466	0.0429418279539109\\
7.36806299728077	0.0530647086375904\\
11.3796204055278	0.0852853377120293\\
17.5752786888082	0.184026825468062\\
27.1441761659491	0.51435098077278\\
41.9228800165354	1.70504946803217\\
64.7478802869525	5.99526347929877\\
100	44.0115534715157\\
};
%\addlegendentry{Est RanNB}

\addplot [color=mycolor3, only marks, mark size=2.0pt, mark=o, mark options={solid, mycolor3}]
  table[row sep=crcr]{%
2	0.0515221733656968\\
3.08890420989276	0.045443072884528\\
4.7706646089466	0.0446376603253829\\
7.36806299728077	0.0541928002651413\\
11.3796204055278	0.0855089915631785\\
17.5752786888082	0.180626231444388\\
27.1441761659491	0.500924472685501\\
41.9228800165354	1.6225396047417\\
64.7478802869525	5.57756078339411\\
100	19.929886543222\\
};
%\addlegendentry{PEB RanNB}
\end{axis}
\node[rotate=40,fill=white] (BOC6) at (5.5cm,2.75cm){\footnotesize Directional};
\node[rotate=40,fill=white] (BOC6) at (5.5cm,4.85cm){\footnotesize Random};
\draw (5.5cm,4.35cm) ellipse (.1cm and .2cm);
\draw (5.5cm,3.15cm) ellipse (.1cm and .2cm);
\node[rotate=0,fill=white] (BOC6) at (4.5cm,5.8cm){\footnotesize c) $M=64^2$, $N=1500$};
\end{tikzpicture}%
    \end{subfigure}  %
    
    \caption{Estimation error and the CRB bounds for \ac{ue} position along the path $[-r/\sqrt{2},r/\sqrt{2},-10]$, where $r$ varies between $2$m to $100$m considering NB and WB models, and directional and random \ac{ris} phase profiles. Results are presented for different combinations of the number of \ac{ris} elements ($M$) and subcarriers ($N$): a) $M=64^2$, $N=3000$,  b) $M=32^2$, $N=3000$,  c) $M=64^2$, $N=1500$.  }
    \label{fig:MN}
\end{figure*}
%--------------------------------

\subsection{Wideband effects}\label{sec:Results_wbEffects}
In this section, we study the accuracy of our estimator in presence of spatial-WB effects using numerical results. To do so, we calculate the PEB and evaluate the UE position estimation error considering the random and directional RIS profiles described in Section\,\ref{sec:beamforming}.
We place the \ac{ue} at $[-r/\sqrt{2} , r/\sqrt{2}, -10]$ for $r\in[2 , 100]$ (in meters). \rev{ Figure\,\ref{fig:struct} demonstrates the placement of BS, RIS, and UE in the 3D space.}
Furthermore,  in Fig.\,\ref{fig:MN} we consider the data transmission through the spatial-WB channel in \eqref{eq:Yb}--\eqref{eq:Yr} and also the spatial-NB one in \eqref{eq:Ybn}--\eqref{eq:Yrn}.  For each point, we average the results over 20  sets of RIS phase profiles, for each of which we consider 20 noise realizations. 

Fig.\,\ref{fig:MN}\,(a) presents the results for \rev{$M = 64^2$} and $N = 3000$. It can be seen that with the NB channel, the estimator attains the PEB at every point. With the WB channel, the estimator has a noticeably larger error compared to the PEB for low values of $r$. The reason is that for low values of $r$ the angle $\alpha$ in \eqref{eq:condSpWB} becomes large and the assumption \eqref{eq:condSpWB} \rev{does} not hold. Therefore, the mismatch between the WB and the NB channels becomes considerable, and the accuracy of the estimator (which is designed based on the NB channel) deteriorates.  Furthermore, one can observe that the PEBs for the WB and NB channel models are almost equal, which shows that the performance degradation can be compensated by adopting a better (and more complex) estimator. \rev{Future research can aim to prove mathematically (via FIM analysis) that the changes of PEB due to user mobility and spatial WB effects are indeed negligible.}

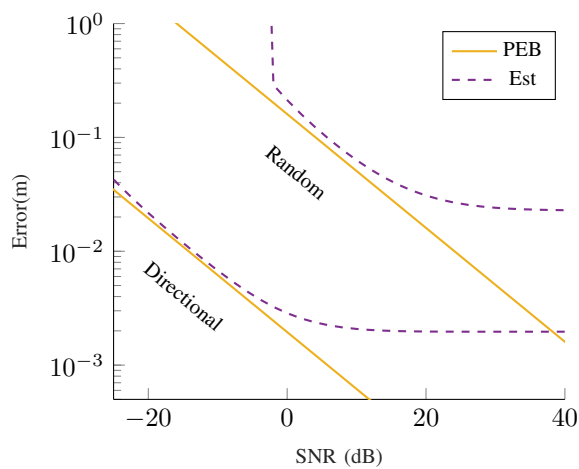
\begin{figure}
    \centering
\definecolor{mycolor1}{rgb}{0.00000,0.44700,0.74100}%
\definecolor{mycolor2}{rgb}{0.85000,0.32500,0.09800}%
\definecolor{mycolor3}{rgb}{0.92900,0.69400,0.12500}%
\definecolor{mycolor4}{rgb}{0.49400,0.18400,0.55600}%

\begin{tikzpicture}

\begin{axis}[%
width=6cm,
height=5cm,
at={(1.011in,0.719in)},
scale only axis,
xmin=-25,
xmax=40,
xlabel style={font=\color{white!15!black}},
xlabel={\footnotesize SNR (dB)},
ymode=log,
ymin=0.0005,
ymax=1,
yminorticks=true,
ylabel style={font=\color{white!15!black}},
ylabel={\footnotesize Error(m)},
axis background/.style={fill=white},
axis x line*=bottom,
axis y line*=left
]
\addplot [color=mycolor3, line width=.8pt]
  table[row sep=crcr]{%
-27.9776027749778	0.0488668355665843\\
-25.9776027749777	0.0388160728972267\\
-23.9776027749777	0.0308327026633858\\
-21.9776027749777	0.0244912862783771\\
-19.9776027749777	0.0194541201956354\\
-17.9776027749777	0.0154529569531098\\
-15.9776027749777	0.0122747200178316\\
-13.9776027749777	0.0097501566835033\\
-11.9776027749777	0.00774482474669128\\
-9.97760277497774	0.00615193296928851\\
-7.97760277497774	0.00488665405563644\\
-5.97760277497774	0.00388160728972256\\
-3.97760277497774	0.00308327026633838\\
-1.97760277497774	0.0024491286278381\\
0.0223972250222602	0.0019454120195635\\
2.02239722502226	0.00154529569531116\\
4.02239722502226	0.00122747200178357\\
6.02239722502226	0.000975015668350108\\
8.02239722502226	0.000774482474669217\\
10.0223972250223	0.000615193296928764\\
12.0223972250223	0.000488665405563701\\
14.0223972250223	0.000388160728972205\\
16.0223972250223	0.000308327026633803\\
18.0223972250223	0.000244912862783813\\
20.0223972250223	0.000194541201956356\\
22.0223972250223	0.000154529569531122\\
24.0223972250223	0.000122747200178361\\
26.0223972250223	9.75015668350089e-05\\
28.0223972250223	7.74482474669022e-05\\
30.0223972250223	6.15193296928788e-05\\
32.0223972250223	4.8866540556379e-05\\
34.0223972250223	3.88160728972199e-05\\
36.0223972250223	3.08327026633817e-05\\
38.0223972250223	2.44912862783821e-05\\
40.0223972250223	1.94541201956355e-05\\
42.0223972250223	1.54529569531096e-05\\
44.0223972250223	1.22747200178376e-05\\
46.0223972250223	9.75015668350175e-06\\
};
\addplot [color=mycolor4, dashed, line width=.8pt]
  table[row sep=crcr]{%
-27.9776027749778	0.0690146291267188\\
-25.9776027749777	0.0493645247767263\\
-23.9776027749777	0.0368698850203365\\
-21.9776027749777	0.0280553514540159\\
-19.9776027749777	0.0216585717207259\\
-17.9776027749777	0.016904870808774\\
-15.9776027749777	0.0132984964690584\\
-13.9776027749777	0.0105356961045713\\
-11.9776027749777	0.00840479390216966\\
-9.97760277497774	0.00675554855599915\\
-7.97760277497774	0.00548947714572607\\
-5.97760277497774	0.00452112476210913\\
-3.97760277497774	0.0037908735334985\\
-1.97760277497774	0.00324946362432963\\
0.0223972250222602	0.00285476207694582\\
2.02239722502226	0.0025724052268312\\
4.02239722502226	0.00237397754158036\\
6.02239722502226	0.00223604782677119\\
8.02239722502226	0.00214169229651004\\
10.0223972250223	0.00207758216088308\\
12.0223972250223	0.00203494356270846\\
14.0223972250223	0.0020071960876546\\
16.0223972250223	0.00199002104798033\\
18.0223972250223	0.00197940654744416\\
20.0223972250223	0.00197272510693079\\
22.0223972250223	0.00196876526190546\\
24.0223972250223	0.00196669081895127\\
26.0223972250223	0.00196585244658713\\
28.0223972250223	0.0019654016632555\\
30.0223972250223	0.00196530315971007\\
32.0223972250223	0.00196538642031248\\
34.0223972250223	0.00196554769522631\\
36.0223972250223	0.00196573873455344\\
38.0223972250223	0.00196592391672132\\
40.0223972250223	0.00196608646033091\\
42.0223972250223	0.00196624933458126\\
44.0223972250223	0.00196638034252114\\
46.0223972250223	0.00196649075658189\\
};\addlegendentry{\footnotesize PEB}
\addplot [color=mycolor3, line width=.8pt]
table[row sep=crcr]{%
-27.9776027749778	4.01615558877276\\
-25.9776027749777	3.19014488769871\\
-23.9776027749777	2.53402215715675\\
-21.9776027749777	2.01284534682247\\
-19.9776027749777	1.59885989111005\\
-17.9776027749777	1.27001955488039\\
-15.9776027749777	1.00881239109357\\
-13.9776027749777	0.801328165794168\\
-11.9776027749777	0.636517587364976\\
-9.97760277497774	0.505603891548291\\
-7.97760277497774	0.401615446643838\\
-5.97760277497774	0.319014488768303\\
-3.97760277497774	0.253402215713755\\
-1.97760277497774	0.201284534685019\\
0.0223972250222602	0.15988598911087\\
2.02239722502226	0.127001955488979\\
4.02239722502226	0.100881239109771\\
6.02239722502226	0.0801328165780548\\
8.02239722502226	0.0636517587365793\\
10.0223972250223	0.0505603891548468\\
12.0223972250223	0.0401615446639582\\
14.0223972250223	0.0319014488769437\\
16.0223972250223	0.0253402215712677\\
18.0223972250223	0.0201284534681734\\
20.0223972250223	0.0159885989114524\\
22.0223972250223	0.0127001955489701\\
24.0223972250223	0.0100881239110201\\
26.0223972250223	0.00801328165783641\\
28.0223972250223	0.0063651758736997\\
30.0223972250223	0.00505603891543899\\
32.0223972250223	0.00401615446641646\\
34.0223972250223	0.00319014488764421\\
36.0223972250223	0.00253402215713843\\
38.0223972250223	0.00201284534684627\\
40.0223972250223	0.00159885989112108\\
42.0223972250223	0.00127001955488797\\
44.0223972250223	0.00100881239110049\\
46.0223972250223	0.00080132816579588\\
};\addlegendentry{\footnotesize Est}
\addplot [color=mycolor4, dashed, line width=.8pt, forget plot]
  table[row sep=crcr]{%
-27.9776027749778	293331.697210795\\
-25.9776027749777	205027.379916116\\
-23.9776027749777	191074.363875018\\
-21.9776027749777	245351.533756499\\
-19.9776027749777	186493.980503031\\
-17.9776027749777	205443.721549124\\
-15.9776027749777	269951.846271619\\
-13.9776027749777	237127.496978865\\
-11.9776027749777	253703.303392351\\
-9.97760277497774	171151.260759388\\
-7.97760277497774	215661.153090663\\
-5.97760277497774	80628.331379302\\
-3.97760277497774	2368.3320984873\\
-1.97760277497774	0.28900675104512\\
0.0223972250222602	0.21229675068283\\
2.02239722502226	0.159942672576353\\
4.02239722502226	0.123190940607629\\
6.02239722502226	0.0967440486408299\\
8.02239722502226	0.0773647545709812\\
10.0223972250223	0.0630042344588563\\
12.0223972250223	0.0523079062132108\\
14.0223972250223	0.0443332106896089\\
16.0223972250223	0.0384245249679901\\
18.0223972250223	0.0340504511629495\\
20.0223972250223	0.0308414108211386\\
22.0223972250223	0.0284999573307896\\
24.0223972250223	0.0268000121804556\\
26.0223972250223	0.0255782919305194\\
28.0223972250223	0.0247075139864008\\
30.0223972250223	0.0240957826587133\\
32.0223972250223	0.0236697768085216\\
34.0223972250223	0.0233754302071771\\
36.0223972250223	0.0231756922222045\\
38.0223972250223	0.0230404431930854\\
40.0223972250223	0.0229497305999058\\
42.0223972250223	0.0228892452754484\\
44.0223972250223	0.0228494317230224\\
46.0223972250223	0.0228232479995575\\
};
\end{axis}
\node[rotate=-40,fill=white] (BOC6) at (3.5cm,3.2cm){\footnotesize Directional};
\node[rotate=-40,fill=white] (BOC6) at (5cm,4.85cm){\footnotesize Random};

\end{tikzpicture}%
    \caption{\rev{Position error for the UE position $[-5/\sqrt{2}, -5/\sqrt{2}, -10]$  for directional and random RIS phase profiles vs the received SNR (of the direct path). }}
    \label{fig:snr}
\end{figure}

%-------------  figure---------------
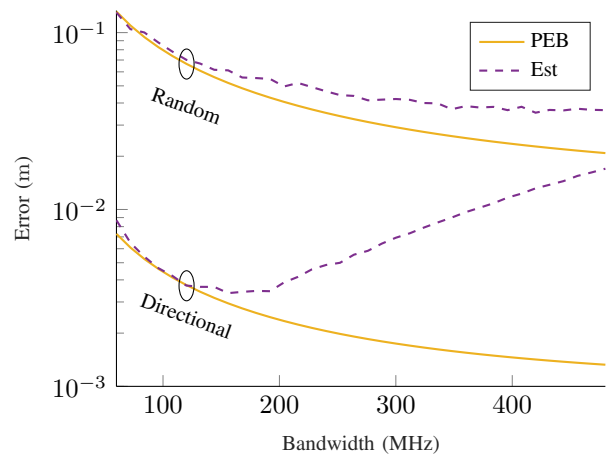
\begin{figure}[t]
    \centering
\definecolor{mycolor1}{rgb}{0.00000,0.44700,0.74100}%
\definecolor{mycolor2}{rgb}{0.85000,0.32500,0.09800}%
\definecolor{mycolor3}{rgb}{0.92900,0.69400,0.12500}%
\definecolor{mycolor4}{rgb}{0.49400,0.18400,0.55600}%
\begin{tikzpicture}

\begin{axis}[%
width=6.5cm,
height=5cm,
at={(1.011in,0.733in)},
scale only axis,
xmin=60,
xmax=480,
xlabel style={font=\color{white!15!black}},
xlabel={\footnotesize Bandwidth (MHz)},
ymode=log,
ymin=0.001,
ymax=0.133269248183589,
yminorticks=true,
ylabel style={font=\color{white!15!black}},
ylabel={\footnotesize  Error ($\mr{m}$)},
axis background/.style={fill=white},
axis x line*=bottom,
axis y line*=left,
legend style={legend cell align=left, align=left, draw=white!15!black},
legend pos= north east]
\addplot [color=mycolor3, line width=.8pt]
  table[row sep=crcr]{%
60	0.00728298202700852\\
72	0.0060925602195724\\
84	0.00524734071365655\\
96	0.00461650884979387\\
108	0.00412856671777917\\
120	0.00374068961448338\\
132	0.00342557862672393\\
144	0.00316494853038141\\
156	0.00294611929048598\\
168	0.00276008091611041\\
180	0.00260026666309137\\
192	0.00246173887631412\\
204	0.00234068301992314\\
216	0.00223411974048295\\
228	0.0021397198483126\\
240	0.00205564266789529\\
252	0.00198039323363029\\
264	0.0019127280256735\\
276	0.00185161308949154\\
288	0.00179620043202495\\
300	0.00174579155516276\\
312	0.00169979283580584\\
324	0.00165768657234661\\
336	0.00161902540670894\\
348	0.00158343269980619\\
360	0.00155059097208031\\
372	0.00152022200389428\\
384	0.00149207461708112\\
396	0.00146592604237471\\
408	0.00144158607129788\\
420	0.00141889232120744\\
432	0.00139769913162165\\
444	0.0013778712264149\\
456	0.00135928638317199\\
468	0.00134183961905303\\
480	0.00132544074323171\\
};
\addlegendentry{\footnotesize PEB }

\addplot [color=mycolor4, dashed, line width=.8pt]
  table[row sep=crcr]{%
60	0.00869018887147358\\
72	0.00648375899138945\\
84	0.00540970579781964\\
96	0.00466092575138127\\
108	0.00421645341197352\\
120	0.00371601566159195\\
132	0.00365282101460005\\
144	0.00365049225057509\\
156	0.00336605048308519\\
168	0.00342119175451991\\
180	0.00346499035237548\\
192	0.00345167301569467\\
204	0.00387070646190705\\
216	0.00416162577788166\\
228	0.00459107143294658\\
240	0.00485545809236418\\
252	0.00499498137571398\\
264	0.00555186903969136\\
276	0.00583916788249351\\
288	0.00641504338241811\\
300	0.00692279688165847\\
312	0.00733677587971589\\
324	0.00789958755599314\\
336	0.00844539134139985\\
348	0.00900338873577514\\
360	0.00965750567346425\\
372	0.0102335563469538\\
384	0.0108484709563146\\
396	0.0116727275057593\\
408	0.0122977216615652\\
420	0.0131668559517847\\
432	0.013793278175346\\
444	0.0144672906536469\\
456	0.015479683471925\\
468	0.0161900616765057\\
480	0.0170037686537863\\
};
\addlegendentry{\footnotesize Est }

\addplot [color=mycolor3, line width=.8pt]
  table[row sep=crcr]{%
60	0.131406875162968\\
72	0.109738654003769\\
84	0.0942978083676878\\
96	0.0827482241230271\\
108	0.0737924358768889\\
120	0.0666521406254194\\
132	0.0608320988401614\\
144	0.0560020284654732\\
156	0.0519331473593121\\
168	0.0484620362977067\\
180	0.0454689327946614\\
192	0.0428640480991146\\
204	0.0405786664696852\\
216	0.0385592762062289\\
228	0.036763627232349\\
240	0.0351579639477083\\
252	0.0337149826135532\\
264	0.0324123152433946\\
276	0.0312314463359676\\
288	0.0301569460422091\\
300	0.0291758831277437\\
312	0.0282773314887638\\
324	0.0274519668261998\\
336	0.0266917744115911\\
348	0.025989847360166\\
360	0.0253402215712677\\
372	0.0247377140185006\\
384	0.0241777787919829\\
396	0.0236564070623504\\
408	0.0231700653398302\\
420	0.0227156407241893\\
432	0.0222903747287062\\
444	0.0218917996884229\\
456	0.0215176987636234\\
468	0.0211660867231947\\
480	0.0208351899806514\\
};
%\addlegendentry{PEB RanWB}

\addplot [color=mycolor4, dashed, line width=.8pt]
  table[row sep=crcr]{%
60	0.129990642737061\\
72	0.104622453217469\\
84	0.10016476212482\\
96	0.0879051775648337\\
108	0.0781524028635345\\
120	0.0702078349848863\\
132	0.0667890721461467\\
144	0.0616403189938853\\
156	0.0611813425787199\\
168	0.0557573647759513\\
180	0.0553378455399396\\
192	0.0547050202459727\\
204	0.0495944905596311\\
216	0.0519250263764981\\
228	0.0491706653209972\\
240	0.0465863621432509\\
252	0.0443283140704365\\
264	0.0438542209571184\\
276	0.0414273893965414\\
288	0.0417857794893256\\
300	0.0421335290510471\\
312	0.0416835870957357\\
324	0.0399980000190409\\
336	0.0398073249282632\\
348	0.0368786990299114\\
360	0.0384245260782664\\
372	0.037728317502282\\
384	0.0380333537345431\\
396	0.0364394512299973\\
408	0.038082608121807\\
420	0.0352574983135581\\
432	0.0365533568324639\\
444	0.0363216687115616\\
456	0.0371178522640913\\
468	0.0366074393460168\\
480	0.0364489701279085\\
};
%\addlegendentry{Est RanWB}
\end{axis}
\node[rotate=-20,fill=white] (BOC6) at (3.5cm,2.8cm){\footnotesize Directional};
\node[rotate=-20,fill=white] (BOC6) at (3.5cm,5.6cm){\footnotesize Random};
\draw (3.5cm,3.2cm) ellipse (.1cm and .2cm);
\draw (3.5cm,6.15cm) ellipse (.1cm and .2cm);
\end{tikzpicture}%
    \caption{Estimation error and CRB bounds for the \ac{ue} position at $[-5/\sqrt{2}, 5/\sqrt{2}, -10]$ as a function of signal bandwidth ($B$). Directional and random phase profiles were considered.   }
    \label{fig:freqs}
\end{figure}
%--------------------------------

In Fig.\,\ref{fig:MN}\,(a), it  can  be seen that \rev{the estimation error due to spatial-WB effects} has a more pronounced effect for directional beamforming than the random one, which is mainly due to higher SNRs in the directional case which makes the influence of the distortion caused by the spatial-WB effects more pronounced. Fig.\,\ref{fig:MN}\,(b) and Fig.\,\ref{fig:MN}\,(c), present the same results for a system with half of the RIS size and half of the bandwidth of the system considered in Fig.\,\ref{fig:MN}\,(a). Apart from a natural degradation in localization accuracy, it can be seen that the WB effects diminish. This can be justified by \eqref{eq:condSpWB}. Also, we note that for large values of $r$ and random beamforming the estimation error in Fig.\,\ref{fig:MN}\,(b) cannot follow the PEB due to low values of SNR.

%-------------  figure Mobility with cdfs---------------
\begin{figure*}[t]
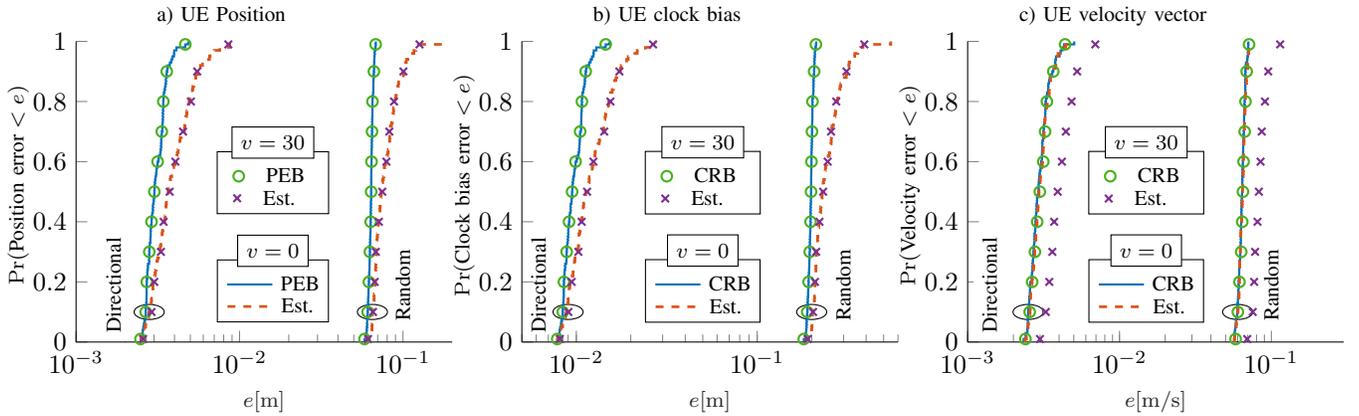

    \centering
    \begin{subfigure}[b]{0.32\textwidth}
    \input{Fig7a}
    \end{subfigure}
    %\hspace{1cm}
    \begin{subfigure}[b]{0.32\textwidth}
    \input{Fig7b}
    \end{subfigure}
    \begin{subfigure}[b]{0.32\textwidth}
    \input{Fig7c}
    \end{subfigure}  %
    
    \caption{ CDF of estimation error and CRB bounds for $100$ realizations of directional and random \ac{ris} phase profiles for a) UE position, b) UE clock bias, and c) UE velocity $\bm{v}$. The \ac{ue} has the position $[-10, 10, -10]$ and velocity $[-v, v, 0]$, where $v\in\{0,30\}$m/s. }
    \label{fig:mobilityCDFs}
\end{figure*}
%--------------------------------

\rev{In Fig.\,\ref{fig:snr}, we demonstrate the position error for the UE location at $[-5/\sqrt{2}, -5/\sqrt{2}, -10]$ for a large range of the received SNR of the direct path. It can be seen that for SNRs lower than $0$ dBm, the estimator fails to estimate the UE location for random RIS profiles. Also, it can be seen that at high SNRs the estimation error saturates for both directional and random phase profiles due to the spatial WB effects. Based on our simulation results (not included in this paper), similar behavior can also be observed for the estimation error of the AOD.}

To study the WB effects more closely in Fig.\,\ref{fig:freqs}, we present the PEB and the estimation errors at $r=5$ for a large range of signal bandwidth ($B=N\deltaf$). As can be seen, the PEBs decrease with $B$, which shows that a better localization performance can be attained with higher bandwidths. However, our estimator, which is designed based on the NB model, does not show such behavior. Specifically with the directional beamforming, after $B = 140$ MHz the distortion caused by the WB effects causes a higher positioning error.

\subsection{Mobility effects}\label{sec:Results_mobilityEffects}
Fig.\,\ref{fig:mobilityCDFs} presents the \acp{cdf} of the estimation error and the \ac{crb} for $100$ different realizations of random and directional RIS phase profiles. For each RIS phase profile we generated $1000$ noise realizations to accurately calculate the estimation error. We consider two \ac{ue} velocities: One where \ac{ue} is static and one where the \ac{ue} velocity vector is set to \rev{$\bm{v}=[-30, 30, 0]\, \mathrm{m/s}$}. We consider the estimation of the UE position in Fig.\,\ref{fig:mobilityCDFs}\,(a), UE clock bias in Fig.\,\ref{fig:mobilityCDFs}\,(b), and UE radial velocity vector $v_{\mathrm{r}}$ in Fig.\,\ref{fig:mobilityCDFs}\,(c).  The velocity vector is estimated based on the radial velocities and the relations \eqref{eq:vub} and \eqref{eq:vur} and by assuming that the estimator has the prior knowledge that $[\bm{v}]_3=0$.

It can be seen from Fig.\,\ref{fig:mobilityCDFs} that in addition to the UE position, the \ac{ue} velocity vector and also the UE clock bias can be estimated. Therefore, the \ac{ue} can be synchronized to the BS. There is a small reduction in the accuracy of velocity estimation for the high-mobility user compared to the static one. This is due to the error in position estimation which causes error in the estimation of $\bm{v}$ from the estimated radial velocities, $\hat{v}_{\mathrm{b}}$ and $\hat{v}_{\mathrm{r}}$. Apart from this, it is apparent that the \ac{ue} velocity does not affect the estimation accuracy, both in terms of analytical bounds and also estimation error.  This can be explained based on the fact that the \ac{ue} radial velocities can be estimated \rev{with the accuracy of up to $0.1\,\mathrm{m/s}$} and then their effects can be removed from the received signal.

%-------------  figure---------------
\begin{figure}[t]
    \centering
    \input{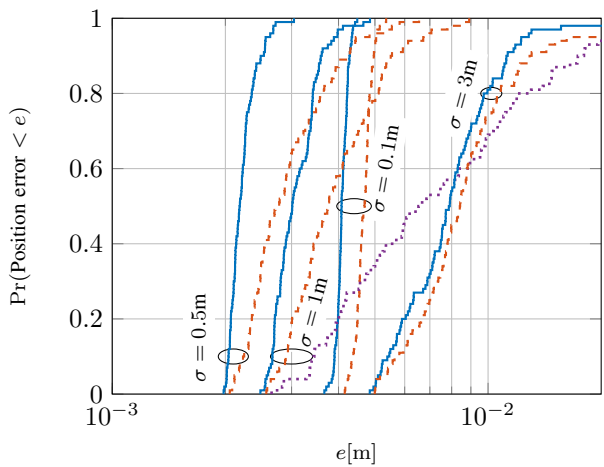}
    \caption{CDF of estimation error (dashed lines) and CRB bounds (solid lines) for $100$ realizations of directional  \ac{ris} phase profiles constructed based on different uncertainty radii ($\sigma$). The estimation error for $\sigma=1$m in presence of $10$ scatterers is also shown (the dotted line). The \ac{ue} has the position $[-10, 10, -10]$. }
    \label{fig:sigma}
\end{figure}
%--------------------------------

\subsection{Uncertainty radius and scatterers}\label{sec:radius}
Fig.\,\ref{fig:sigma} demonstrates the \ac{cdf} of the position error for $100$ realizations of the directional RIS beams for different values of $\sigma$. It can be seen that the optimal performance among the considered values of $\sigma$ is obtained by $\sigma=0.5$\,m. For very small values of $\sigma$ (like $\sigma = 0.1$\,m), all the transmitted beams become almost similar to each other and therefore accurate \ac{aod} estimation cannot be performed due to lack of beam diversity. Furthermore, with larger values of $\sigma$ there is a probability that none of the transmitted beams hits the \ac{ue}, which results in low SNR and high estimation error. This is the reason why the \ac{cdf} of the estimation error becomes saturated around $0.95$ for $\sigma=3$m. Furthermore, we examine the performance of our estimator in presence of $10$ scatterers, whose radio cross sections are equal to $0.1$ $\text{m}^2$ and are distributed randomly on a disc placed on the $z=-11$ plane, centered at $[0,0,-11]$ with $5$ meters radius. The channel gains for scatterers are calculated based on the radar range equation (see e.g., \cite[Eq.\,(23)]{ellingson2019path}). It can be seen that although the presence of the scatterers can degrade the localization accuracy, it is still possible to perform cm-level positioning. 

\section{Conclusion}\label{sec:conclusion}
We analyzed the influence of UE mobility and spatial-WB effects on the accuracy of an RIS-enabled SISO system by deriving \ac{crb} and also devising an estimator. Based on our numerical results, it was shown that the UE mobility does not have any notable effects on the estimation accuracy of the \ac{ue} state. This was shown in Fig.\,\ref{fig:mobilityCDFs}, where both the  bounds and the estimation errors are virtually equal for a static \ac{ue} and a \ac{ue} with a very high speed. Our proposed estimator dealt with the \ac{ue} mobility by successively estimating the radial velocities and compensating for their effects. Our results suggest that the studies that assume static users can be potentially extended to account for the \ac{ue} mobility without any significant performance degradation. With regard to spatial-WB effects it was shown that these effects do not change the analytical bounds and therefore the performance of the optimal estimator. However, for a low-complexity estimator that ignores the spatial-WB effects (such as the one presented in this work), they can degrade the localization accuracy, especially for large signal bandwidth and RIS sizes. Specifically, it was shown in Fig.\ref{fig:freqs} that for some typical system values increasing the number of subcarriers can decrease the estimation accuracy indicating the existence of an optimal signal bandwidth. This result shows the importance of devising a low-complexity estimator that can cope with the spatial-WB effects in  dynamic systems, which is an interesting topic for future research.

\begin{appendices}

 \section{Spatial-wideband model under UE mobility} \label{app:Specially_wideband_Ch_Model}
 In this appendix, we derive the received signal coming through the reflected path ($\bm{Y}_{\mathrm{r}}$) by taking into account spatial-WB effects \rev{\cite{wang2018spatial,dovelosintelligent,face_squint,chen2021beam}} and UE mobility \rev{\cite{Matthiesen_continuous,basar2019reconfigurable,huang_icc21,sun_wcl_doppler}}. The received signal from the direct path ($\bm{Y}_{\mathrm{b}}$) in \eqref{eq:Yb} can be derived in the same fashion. We use the same notation as in Section~\ref{sec:systemModChannelModel}. \rev{In addition, for the derivations, we adopt an approach similar to those in \cite{face_squint,bjornson2021reconfigurable,wang2018spatial}, where we compute the total path delay between the BS and the UE, including the BS-to-RIS delay, the (adjustable) delay at the RIS and the RIS-to-UE delay, along with the Doppler effects due to UE mobility.}
 
 \subsection{Transmit signal model}
The transmitted OFDM  baseband  signal can be expressed as
\begin{align}\label{eq_ofdm_baseband_all}
    s(t) = \sum_{\ell=0}^{L-1} \sll(t),
\end{align}
where
\begin{equation}\label{eq_ofdm_baseband}
\sll(t) = \frac{1}{\sqrt{N}} \sum_{n = 0}^{N-1}  \xnl \, e^{\jmath 2 \pi n \deltaf t} \rect{\frac{t - \ell \Tsym}{\Tsym}} 
\end{equation} 
denotes the OFDM signal for the $\thn{\ell}$ symbol, $\xnl$ is the complex pilot symbol on the $\thn{n}$ subcarrier for the $\thn{\ell}$ symbol, and $\rect{t}$ is a rectangular function that takes the value $1$ for $t \in [0, 1 )$ and $0$ otherwise. Then, the upconverted transmit signal can be written as
\begin{equation}\label{eq_passband_st}
\stilde(t) = \Re \left\{  s(t) e^{j  2 \pi \fc t } \right\}.
\end{equation}

\subsection{Receive signal model}
Based on the transmit signal model in \eqref{eq_passband_st}, the passband received signal at the UE due to the reflected path through the RIS is given by \rev{\cite{face_squint}}
\begin{align}\label{eq_rec_passband}
    \ytilde(t) = \Re \left\{ \grtilde \sum_{m=0}^{M-1} s(t- [\Tauv(t)]_m) e^{\jmath 2 \pi \fc (t-[\Tauv(t)]_m) } \right\}, 
\end{align}
where $\grtilde$ is the complex path gain and the vector $\Tauv(t)\in \mathbb{R}^{M}$ contains the delays between the BS and the UE through the different elements of the RIS. It can be computed as
\begin{align}\label{eq_taumt}
    \Tauv(t) = \Taubr + \Taur(t) + \Tauru(t) + \deltat,
\end{align}
where the vector $\Taubr$ contains the delays between the BS and the elements of the RIS
\begin{align}
    [\Taubr]_m = \frac{ \norm{\pbb - \pbrm} }{c} 
\end{align}
with $\pbrm$ denoting the location of the $\thn{m}$ RIS element, $[\Taur(t)]_m$ denotes the delay incurred by the $\thn{m}$ element of the RIS  at time $t$ \cite{bjornson2021reconfigurable}, $[\Tauru(t)]_m = [\Tauru]_m - \murk t$ represents the time-varying delay \rev{\cite{basar2019reconfigurable}} from the $\thn{m}$ element to the UE with $\murk = \vurk/c$ and $\vurk$ denoting the radial velocity along the RIS-UE direction in \eqref{eq:vur} and
\begin{align}
    [\Tauru]_m = \frac{ \norm{\pbrm - \pb} }{c} 
\end{align}
is the initial delay (at $t=0$).
The complex baseband received signal after downconversion of \eqref{eq_rec_passband} can be written as \rev{\cite{face_squint}}
\begin{align}\label{eq_rec_baseband}
    y_{\mathrm{r}}(t) =  \grtilde \sum_{m=0}^{M-1} s(t-[\Tauv(t)]_m) e^{-\jmath 2 \pi \fc [\Tauv(t)]_m }.
\end{align}
Plugging \eqref{eq_ofdm_baseband_all} and \eqref{eq_ofdm_baseband} into \eqref{eq_rec_baseband}, we have
\begin{align}\label{eq_rec_baseband2}
    y_{\mathrm{r}}(t) &=  \grtilde \sum_{m=0}^{M-1} \sum_{\ell=0}^{L-1} \frac{1}{\sqrt{N}} \sum_{n = 0}^{N-1}  \xnl \, e^{\jmath 2 \pi n \deltaf (t - [\Tauv(t)]_m )} 
    \\ \nonumber &~~~~~~ \times e^{-\jmath 2 \pi \fc [\Tauv(t)]_m } \rect{\frac{t - [\Tauv(t)]_m - \ell \Tsym}{\Tsym}}.
\end{align}
For the $\thn{\ell}$ symbol, we sample $y_{\mathrm{r}}(t)$ in \eqref{eq_rec_baseband2} at $t = \ell \Tsym + \Tcp + \taumin + k \Tso/N$ for $k = 0, \ldots, N-1$ (i.e., we remove the CP and sample the interval corresponding to the elementary OFDM signal), where 
\begin{align}
    \taumin = \min_{m} [\bm{\tau}(0)]_m
\end{align}
is the arrival time of the reflected path with respect to the receiver's clock (which can be detected\footnote{Since the variation of the delays $\Tauv(t)$ across the RIS elements could be much smaller than the delay resolution, the UE can possibly identify a single correlation peak contributed by all the RIS elements, in which case $\taumin$ is set as the location of that peak.}, e.g., via downlink synchronization signals \cite{TS_38211}). Substituting \eqref{eq_taumt} into \eqref{eq_rec_baseband2}, the discrete-time signal for the \rev{$\thn{k}$ sample of the} $\thn{\ell}$ symbol at the receiver becomes
\begin{align}
    \ylkm &= \grtilde \sum_{m=0}^{M-1} \frac{1}{\sqrt{N}} \sum_{n = 0}^{N-1} \Big[  \xnl \, e^{\jmath 2 \pi n \deltaf (\ell \Tsym + \Tcp + \taumin + k \Tso/N )}  \nonumber\\ 
    &~~~~ \times e^{-\jmath 2 \pi n \deltaf (\taumbr + \taumgell + \taumru + \deltat )} \nonumber
    \\  \label{eq_ylk_long}
    &~~~~ \times e^{\jmath 2 \pi n \deltaf \murk (\ell \Tsym + \Tcp + \taumin + k \Tso/N)}
     \\  \nonumber
     &~~~~ \times e^{-\jmath 2 \pi \fc (\taumbr + \taumgell + \taumru + \deltat) }
     \\  \nonumber
     &~~~~ \times e^{\jmath 2 \pi \fc \murk (\ell \Tsym + \Tcp + \taumin + k \Tso/N) } \Big] 
\end{align}
under the assumption that $[\bm{\tau}(0)]_m - \taumin \leq \Tcp$, which holds in practice since the UE is in the far-field of RIS and the RIS delays $\Taur$ are very small compared to propagation delays $\taumbr$ and $\taumru$ \cite{bjornson2021reconfigurable}. In \eqref{eq_ylk_long}, it is assumed that the RIS profile can change across OFDM symbols and $\taumgell = \left[ \Taur(\ell \Tsym + \Tcp) \right]_m$ represents the delay of the $\thn{m}$ element corresponding to the RIS configuration applied for the $\thn{\ell}$ symbol. 

Since the receiver's clock reference can be set to an arbitrary known epoch, we can set $\taumin = 0$. The received signal  in \eqref{eq_ylk_long}  can be written as
\begin{align} 
    \ylkm&= \frac{\grtilde }{\sqrt{N}} e^{\jmath 2\pi f_c \murk(\Tcp + k \Tso/N)} \sum_{n = 0}^{N-1}\xnltilde  e^{\jmath 2 \pi n  k /N } \nonumber\\
 &~~~~ \times e^{\jmath 2 \pi (\fc+n \deltaf) \murk\ell \Tsym}
 e^{\jmath 2\pi n \deltaf \murk (\Tcp + k \Tso/N)}\label{eq_ylk_long2}\\  
    &~~~~ \times \sum_{m=0}^{M-1}e^{-\jmath 2 \pi (\fc+ n \deltaf) (\taumbr + \taumgell + \taumru + \deltat )},\nonumber
\end{align}
where 
\begin{align}\label{eq_xnl_tilde}
    \xnltilde = \xnl e^{\jmath 2\pi n \deltaf (\ell \Tsym +\Tcp)}.
\end{align}
We define the phase shift induced by the delay $\taumgell$ at the center frequency as $\psi_{\ell,m} = 2\pi\fc\taumgell$, which we assume to be less than $2\pi$ (note that the choice of the RIS configuration $\taumgell$ is under the designer's control \rev{\cite{bjornson2021reconfigurable}} and $\taumgell \in [0, 1/\fc)$ will cover all possible phase shifts). 

To \rev{make} \eqref{eq_ylk_long2} \rev{more compact}, \rev{we will now rely on the following approximations/simplifications:}
\begin{enumerate}
    \item \rev{\textit{frequency-narrowband approximation:}}  
\begin{align}\label{eq:phaseApprox}
 \frac{\fc+n\deltaf}{\fc} \psi_{\ell,m} \approx \psi_{\ell,m},
\end{align}
which holds as long as $B/\fc\ll 1$ \rev{(which is satisfied in our simulations according to Table~\ref{table:par} with $B = 360 \, \rm{MHz}$ and $\fc = 30 \, \rm{GHz}$)}.

    \item \textit{far-field approximation\rev{\footnote{\rev{The far-field approximation in \eqref{eq:ff_taubr} (and, similarly, the one in \eqref{eq:ff_tauru}) can be readily derived by observing that, in the far-field regime, the difference between the BS-to-RIS center distance and the BS-to-$m$-th RIS element distance can be written as a function of $\thetab$, the AoA from the BS to RIS, and $[\bm{Q}]_{:,m}$, the position of the $m$-th RIS element relative to the RIS center.}}}:} 
    \begin{align}
    2 \pi (\fc+n \deltaf) (\bm{\tau}_{\mathrm{br}}-\tau_{\mathrm{br}})& \approx -  \bm{k}^\top(\bm{\theta})\bm{Q} \label{eq:ff_taubr}\\
    2 \pi (\fc+n \deltaf) (\bm{\tau}_{\mathrm{ru}}-\tau_{\mathrm{ru}})& \approx - \bm{k}^\top(\bm{\phi})\bm{Q}, \label{eq:ff_tauru}
\end{align}
where $\tau_{\mathrm{br}} = \Vert\bm{p}_{\mathrm{b}}-\bm{p}_{\mathrm{r}}\Vert/c$, $\tau_{\mathrm{ru}} = \Vert\bm{p}_{\mathrm{r}}-\bm{p}\Vert/c$, and $\bm{k}$ and $\bm{Q}$ are defined in \eqref{eq:WaveNumVect} and \eqref{eq:Q}, respectively.

    \item \rev{\textit{negligible phase term under practical velocity values\footnote{The phase of the \ac{lhs} of \eqref{eq:vApprox} can be upper bounded with $2\pi B \Tsym \vurk/c $, which for the values in Table~\ref{table:par} and $\vurk=30\,\mathrm{m}/\mathrm{s}$ is about $2\cdot 10^{-3}$.}:}}
    \begin{align}\label{eq:vApprox}
    e^{\jmath 2\pi n \deltaf \murk (\Tcp + k \Tso/N)}\approx 1.
\end{align}
\end{enumerate}
\rev{In addition}, we define $[\bm{\gamma}_{\ell}]_m = e^{\rev{-}\jmath \psi_{\ell,m}}$ to indicate the RIS phase profile, and a constant phase reference $\psi_{r} \rev{\triangleq} 2\pi f_{\mathrm{c}} \rev{\tau_{\mathrm{r}}}$. Using \rev{\eqref{eq:phaseApprox}--\eqref{eq:ff_tauru}}, the last summation in \eqref{eq_ylk_long2} can be written as
\begin{align} \label{eq_approx_start}
    &\sum_{m=0}^{M-1} e^{-\jmath 2 \pi (\fc+ n \deltaf) (\taumbr + \taumgell + \taumru + \deltat )} 
    \\ \nonumber &= \rev{\sum_{m=0}^{M-1} e^{-\jmath 2 \pi (\fc+ n \deltaf) (\taumbr - \tau_{\mathrm{br}})} e^{-\jmath 2 \pi (\fc+ n \deltaf) (\taumru - \tau_{\mathrm{ru}})}}
    \\ \nonumber &~~\rev{\times
    e^{-\jmath 2 \pi (\fc+ n \deltaf) (\tau_{\mathrm{br}} + \tau_{\mathrm{ru}} +\deltat ) } e^{-\jmath 2 \pi (\fc+ n \deltaf) \taumgell }}
    \\ \nonumber & \approx  \rev{e^{-\jmath 2\pi n\deltaf  \tau_{\mathrm{r}}}  e^{-j \psi_{r}} \sum_{m=0}^{M-1}
    e^{j\bm{k}(\bm{\theta})^\top [\bm{Q}]_{:,m}} e^{j\bm{k}(\bm{\phi})^\top [\bm{Q}]_{:,m}} [\bm{\gamma}_{\ell}]_m}
    \\  \nonumber
    &= \rev{e^{-\jmath 2\pi n\deltaf  \tau_{\mathrm{r}}}  e^{-j \psi_{r}} \sum_{m=0}^{M-1} [\bm{a}(\bm{\theta})]_{m} [\bm{\gamma}_{\ell}]_m [\bm{a}(\bm{\phi})]_{m} }
    \\ \label{eq_approx_main}
    &= \rev{e^{-\jmath 2\pi n\deltaf  \tau_{\mathrm{r}}} e^{-j \psi_{r}} [\bm{A}(\bm{\phi})]_{n,\ell}} ~,
\end{align}
where the matrix $\bm{A}(\bm{\phi})$ is defined in \eqref{eq:aphi} and $\tau_{\mathrm{r}}$ in \eqref{eq:taur}. 

By substituting \eqref{eq_approx_main} and \eqref{eq:vApprox} into \eqref{eq_ylk_long2}, we obtain
\begin{align} 
    \ylkm&=
 \frac{\gr}{\sqrt{N}} e^{\jmath 2\pi  \vurk k \Tso/(\lambda N)} \sum_{n = 0}^{N-1}\xnltilde  e^{\jmath 2 \pi n  k /N } \nonumber\\
 &~~~~ \times e^{\jmath 2 \pi \vurk \ell \Tsym/\lambda_n }\label{eq_ylk_long5} e^{-\jmath 2\pi n\deltaf  \tau_{\mathrm{r}}}  [\bm{A}(\bm{\phi})]_{n,\ell}.
\end{align}
Here, we used $(\fc+n \deltaf)\murk = \vurk(\fc+n \deltaf)/c =\vurk/\lambda_n$, where $\lambda_n$ is defined in \eqref{eq:lambda_n}. Also, we have
\begin{align}
    \gr = \grtilde e^{\jmath 2\pi \fc \murk \Tcp}e^{\rev{-}\jmath \psi_{\mathrm{r}}}.
\end{align}
 Assuming $\xnltilde = 1$ for all\footnote{According to \eqref{eq_xnl_tilde}, the pilot symbols $\xnl$ can be chosen such that $\xnltilde = 1$ for the sake of simplicity of analysis. The signal model can be straightforwardly extended to the case of arbitrary pilot symbols. In addition, the effects of transmit power can be modeled by adjusting the noise variance in \eqref{eq:channelModel:WB}.} $n$ and $\ell$, the summation in \eqref{eq_ylk_long5} can be written via the DFT matrix $\bm{F}$ in \eqref{eq:dft} as
\begin{align}
    \widetilde{\bm{Y}}_{\mathrm{r}} = \gr \bm{E}(\vurk) \bm{F}^{\mathrm{H}}\left(\bm{D}(\tau_{\mathrm{r}})\odot \bm{A}(\bm{\phi})\odot \rev{\ccbigw(\vurk)} \right),
\end{align}
\rev{where $\bm{D}(\tau)$, $\ccbigw(v)$ and $\bm{E}(v)$ are defined, respectively, in \eqref{eq:matrixD}, \eqref{eq:Cmatrix} and \eqref{eq:Ematrix}.}
Finally, we define 
\begin{align}
    \bm{Y}_{\mathrm{r}} = \bm{F} \widetilde{\bm{Y}}_{\mathrm{r}}
\end{align}
to obtain \eqref{eq:Yr}.

%%%%%%%%%%%%%%%%%%%%%%%%%%%%%%%%%%
\section{\rev{Conditions of Validity for Spatial-Narrowband Approximation in \eqref{eq:Ybn}--\eqref{eq:Yrn}}}\label{app_nb_valid}
\rev{In this part, we derive the conditions under which the spatial-narrowband approximation in \eqref{eq:Ybn}--\eqref{eq:Yrn} is valid. To this end, we explore when $\ccbigw(v)$ and $\aabigw(\bm{\phi})$ in the spatial-wideband model \eqref{eq:Yb}--\eqref{eq:Yr} can be approximated as $\ccbig(v)$ and $\aabig(\bm{\phi})$ in the spatial-narrowband model \eqref{eq:Ybn}--\eqref{eq:Yrn}, respectively.}

\rev{\subsection{Condition of Validity for Approximation of \eqref{eq:Cmatrix}}}
\rev{For the transition from $[\ccbigw(v)]_{n,\ell}$ in \eqref{eq:Cmatrix} to $[\ccbig(v)]_{n,\ell}$ in \eqref{eq:CNmatrix} to be valid, the following approximation must hold $\forall \ell, n$:
\begin{align}
    e^{\jmath 2\pi \ell \Tsym  v/\lambda_n} &\approx e^{\jmath 2\pi \ell \Tsym  v/\lambda } ~,
\end{align}
which requires
\begin{subequations}
\begin{align}
    e^{\jmath 2\pi \ell \Tsym  v (\fc + n \deltaf) / c} &\approx e^{\jmath 2\pi \ell \Tsym  v \fc  / c }
    \\ \label{eq_cmatrix_approx_c1}
    e^{\jmath 2\pi \ell \Tsym v n \deltaf /c } &\approx 1
    \\ \label{eq_cmatrix_approx_c2}
    L \Tsym B v & \ll c 
    \\ \label{eq_approx_1_final}
    L N v & \ll c 
    \\ \label{eq_approx_1_final2}
    L N \max\{v_{\mathrm{r}},v_{\mathrm{b}}\} &\ll c ~,
\end{align}
\end{subequations}
where \eqref{eq_cmatrix_approx_c2} is obtained by plugging the worst-case conditions $\ell = L-1$ and $n = N-1$ (in terms of approximation quality) into \eqref{eq_cmatrix_approx_c1} and recalling that $B = N \deltaf$, \eqref{eq_approx_1_final} results from $B \Tsym \approx B T = B / \deltaf = N$ (assuming $\Tcp / T$ is small), and \eqref{eq_approx_1_final2} follows by considering the maximum of direct and reflected path velocities.}

\rev{\subsection{Condition of Validity for Approximation of \eqref{eq:aphi_wb}}}

\rev{Similarly, for the transition from $[\aabigw(\bm{\phi})]_{n,\ell}$ in \eqref{eq:aphi_wb} to $[\aabig(\bm{\phi})]_{n,\ell}$ in \eqref{eq:aphi} to be valid, we need
\begin{align}
    e^{\jmath\bm{k}_n(\bm{\psi})^\top [\bm{Q}]_{:,m}} \approx e^{ \jmath\bm{k}(\bm{\psi})^\top [\bm{Q}]_{:,m} }
\end{align}
for any $n$ and the angles $\bm{\psi} \in \{\thetab, \phib \}$, which represent, respectively, the AoA and AoD for the RIS in \eqref{eq:aphi_wb}. From \eqref{eq:WaveNumVect} and the definition of $\bm{q}_{r,s}$ in Sec.~\ref{sec:systemSetup}, this requires
\begin{subequations}
\begin{align}
     &e^{\jmath  \max(M_1,M_2) d  \sin(\alpha)  2 \pi  / \lambda_n }  \approx e^{\jmath  \max(M_1,M_2) d  \sin(\alpha)  2 \pi  / \lambda }
     \\
     &e^{\jmath  \max(M_1,M_2) d  \sin(\alpha)  2 \pi  (\fc + n \deltaf) / c } \nonumber\\
     &\qquad\qquad\qquad\qquad\approx e^{\jmath  \max(M_1,M_2) d  \sin(\alpha)  2 \pi  \fc  / c }
     \\
     &e^{\jmath  \max(M_1,M_2) d  \sin(\alpha)  2 \pi   n \deltaf / c }  \approx 1
     \\ \label{eq_approx_2_final}
     &\max(M_1,M_2) d   \sin(\alpha)   B \ll c ~,
\end{align}
\end{subequations}
where $\alpha$ denotes the angle between the RIS normal ($[0,1,0]^\top$) and the vector $\bm{k}(\bm{\psi})$\footnote{Note that  $[\bm{Q}]_{:,m}$ is orthogonal to the RIS normal;  therefore, only the component of $\bm{k}(\bm{\psi})$ that is orthogonal to the RIS normal contributes to the value of $\bm{k}(\bm{\psi})^\top [\bm{Q}]_{:,m}$. This component has the norm $\sin(\alpha)$. }, and
\eqref{eq_approx_2_final} follows by considering the worst-case scenario (in terms of approximation quality) $n = N-1$.}

%%%%%%%%%%%%%%%%%%%%%%%%%%%%%%%%%%

\section{Choosing the candidate AoDs}\label{app:fft}
In this section, we explain how we select the \acp{aod} $\bm{\phi}$. For the case with existing prior location information $\bm{\xi}$ (see Section~\ref{sec:dirCodebook}), we choose $N_{\mathrm{\phi}}$ points within the sphere centered at $\bm{\xi}$ with radius $\sigma$ (similarly as in Section~\ref{sec:dirCodebook}). Then the set $\{\bm{\phi}_s\}_{s=0}^{N_{\mathrm{\phi}}-1}$ is calculated as the angles from the \ac{ris} towards these points. Furthermore, with directional beams in \eqref{eq:GammaFunction} the calculation of $\bm{z}_s$ in Line\,\ref{CoarseVA_Line4} of Algorithm\,\ref{alg:coarse_vA_phi} can be performed in closed-form (the \ac{rhs} of Line\,\ref{CoarseVA_Line4} reduces to a geometric sum), which reduces the complexity of Algorithm\,\ref{alg:coarse_vA_phi}. 

In the absence of any prior information about the user, the values of $\bm{z}_s$ can be calculated offline since the beams can be set prior to the localization procedure. Furthermore, to reduce the complexity of calculating $\bm{z}_s$, we use 2D \ac{ifft}, which is explained as follows.  We re-write the vector $\bm{a}(\bm{\psi})$ in \eqref{eq:aVector} as 
\begin{align}
    \bm{a}(\bm{\psi}) =  \bm{a}_1(\bm{\psi}) \otimes \bm{a}_2(\bm{\psi}),\label{eq:kronProd}
\end{align}
where 
\begin{align}
     \bm{a}_1(\bm{\psi}) &= e^{\jmath \beta_1}[1,e^{\jmath [\bm{k}(\bm{\psi})]_1d},\dots,,e^{\jmath [\bm{k}(\bm{\psi})]_1 (M_1-1)d}]\\
      \bm{a}_2(\bm{\psi}) &= e^{\jmath \beta_2}[1,e^{\jmath [\bm{k}(\bm{\psi})]_3d},\dots,,e^{\jmath [\bm{k}(\bm{\psi})]_3 (M_2-1)d}],
\end{align}
where $\beta_1 = [\bm{k}(\bm{\psi})]_1(M_1-1)d/2$ and $\beta_2 = [\bm{k}(\bm{\psi})]_3(M_2-1)d/2$. Next, from Line\,\ref{CoarseVA_Line4} we have that
\begin{align}
    [\bm{z}_s]_{k} &= \bm{a}(\theta)^{\top} \diag(\bm{b}_{k}) \bm{a}(\bm{\phi}_s)\\
    &= \bm{a}(\bm{\phi}_s)^{\top}\left( \bm{a}(\theta)\odot \bm{b}_{k} \right)\\
    &= e^{\jmath(\beta_1+\beta_2)}\bm{a}_1(\bm{\phi}_s)^\top \bm{C}_k \bm{a}_2(\bm{\phi}_s),\label{eq:fftMotviation}
\end{align}
where 
\begin{align}
    \bm{C}_k = \left(\bm{a}_1(\theta)\bm{a}_2(\theta)^{\top} \right)\odot \bm{B}_k
\end{align}
and \eqref{eq:fftMotviation} follows from \eqref{eq:kronProd} and the properties of the Kronecker product (see \cite[Eq.~(520)]{MatCookBook}). Motivated by \eqref{eq:kronProd}, we set $\bm{z}_s$  to be the $s$th row of matrix $\bm{Z}_{\mathrm{f}}= [\bm{z}_{\mathrm{f},0},\dots, \bm{z}_{\mathrm{f},L/2-1}]$, where
\begin{align}
    \bm{z}_{\mathrm{f},k} &= \mathrm{vec}\left(\bm{F}_{\mathrm{\phi},1}^\top \bm{C}_k \bm{F}_{\mathrm{\phi},2}\right).\label{eq:fftReal}
\end{align}
Here, $\bm{F}_{\mathrm{\phi},1}\in \mathbb{C}^{M_1\times N_{\mathrm{\phi},1}}$ and $\bm{F}_{\mathrm{\phi},2}\in \mathbb{C}^{M_2\times N_{\mathrm{\phi},1}}$ are IDFT matrices, where $N_{\mathrm{\phi},1}$ and $N_{\mathrm{\phi},2}$ are  design parameters. Furthermore, the \ac{rhs} of \eqref{eq:fftReal} can be calculated using 2D IFFT. The set $\{\bm{\phi_s}\}$ can be calculated as $\{\bm{\phi}_{0,0}, \bm{\phi}_{1,0}, \dots, \bm{\phi}_{N_{\mathrm{\phi},1}-1,N_{\mathrm{\phi},2}-1} \}$, where 
\begin{align}
    [\bm{\phi}_{n_1,n_2}]_{\mathrm{az}} &=\mathrm{atan2}\left(k_2(n_1,n_2),k_1(n_1,n_2)\right)\\
    [\bm{\phi}_{n_1,n_2}]_{\mathrm{el}} &=\mathrm{acos}\left(k_3(n_1,n_2)\right).
\end{align}
Here, 
\begin{align}
    k_1(n_1,n_2)&= f_{\mathrm{r}}\!\!\left(\frac{\lambda n_1}{d N_{\mathrm{\phi},1}}\right)\\
    k_3(n_1,n_2)&=f_{\mathrm{r}}\!\!\left(\frac{\lambda n_2}{d N_{\mathrm{\phi},2}}\right)\\
    k_2(n_1,n_2)&=\sqrt{1-k_1^2-k_3^2},
\end{align}
where the function
$f_{\mathrm{r}} = x-2\lfloor x/2 \rfloor$ compensates for the wrap-around effects. Furthermore, for the values of $n_1$ and $n_2$ if $ k_2(n_1,n_2)$ becomes imaginary  $\bm{\phi}_{n_1,n_2}$ is undefined and the estimator can remove these values from the sets $\{\bm{z}_s\}$ and $\{\bm{\phi}_s\}$.

\end{appendices}

\balance
\bibliographystyle{IEEEtran}
\bibliography{references}
\end{document}